\documentclass[a4paper,11pt]{article}
\pdfoutput=1 

\usepackage{jinstpub} 
\usepackage{caption}
\usepackage{booktabs}
\usepackage{multirow}
\usepackage{multicol}
\usepackage{threeparttable}
\usepackage{subcaption}
\usepackage{graphicx}
\usepackage{siunitx}
\usepackage[symbol]{footmisc}

\newcommand\isotope[2]{\textsuperscript{#2}#1}

\title{Development of a $^{127}$Xe calibration source for nEXO}

\author[1]{ B.~G.~Lenardo}
\author[1]{ C.~A.~Hardy}
\author[2]{ R.~H.~M.~Tsang}
\author[3]{ J.~C.~Nzobadila Ondze}
\author[2]{ A.~Piepke}
\author[3]{ S.~Triambak}
\author[4]{ A.~Jamil}
\author[5]{ G.~Adhikari}
\author[6]{ S.~Al Kharusi}
\author[1]{ E.~Angelico}
% ...

\affiliation[1]{Physics Department, Stanford University, Stanford, CA 94305, USA }
\affiliation[2]{Department of Physics and Astronomy, University of Alabama, Tuscaloosa, AL 35405, USA }
\affiliation[3]{Department of Physics and Astronomy, University of the Western Cape, P/B X17 Bellville 7535, South Africa }
\affiliation[4]{Wright Laboratory, Department of Physics, Yale University, New Haven, CT 06511, USA }
\affiliation[5]{Physics Department, University of California, San Diego, CA 92093, USA }
\affiliation[6]{Physics Department, McGill University, Montr\'eal, Qu\'ebec H3A 2T8, Canada }
% ...

\emailAdd{blenardo@stanford.edu}

\abstract{We study a possible calibration technique for the nEXO experiment using a \isotope{Xe}{127} electron capture source. nEXO is a next-generation search for neutrinoless double beta decay ($0\nu\beta\beta$) that will use a 5-tonne, monolithic liquid xenon time projection chamber (TPC). The xenon, used both as source and detection medium, will be enriched to 90\% in \isotope{Xe}{136}. To optimize the event reconstruction and energy resolution, calibrations are needed to map the position- and time-dependent detector response. The \SI{36.3}{day} half-life of \isotope{Xe}{127} and its small $Q$-value compared to that of \isotope{Xe}{136} $0\nu\beta\beta$ would allow a small activity to be maintained continuously in the detector during normal operations without introducing additional backgrounds, thereby enabling \emph{in-situ} calibration and monitoring of the detector response. In this work we describe a process for producing the source and preliminary experimental tests. We then use simulations to project the precision with which such a source could calibrate spatial corrections to the light and charge response of the nEXO TPC.}


\collaboration[c]{\\(The nEXO Collaboration)}

\begin{document}
\maketitle
\flushbottom

\section{Introduction}
\label{sec:intro}
The nEXO experiment is a planned tonne-scale search for neutrinoless double beta decay ($0\nu\beta\beta$) in $^{136}$Xe using a cylindrical liquid-phase time projection chamber (TPC)~\cite{nEXO_pCDR}.  
The low intrinsic background, 3-dimensional event reconstruction, and powerful self-shielding of the active liquid xenon volume will enable nEXO to achieve the ultra-low backgrounds needed to reach a sensitivity to $0\nu\beta\beta$ beyond a half-life of $10^{28}$~years~\cite{Adhikari_2021}. The TPC provides a dual-channel measurement of interactions in the liquid xenon target: a scintillation signal, detected promptly via photosensors around the barrel of the detector; and an ionization signal, detected by applying a uniform electric field across the TPC to drift the charge to a collection plane at the anode. The two signals provide complementary information that can be combined to enable the reconstruction of both the 3-dimensional position of each interaction vertex and the deposited energy. One of the performance targets for nEXO is to achieve an energy resolution  better than $\sigma / E = 1$\% at the $0\nu\beta\beta$ Q-value (2.457~MeV), which contributes to the rejection of backgrounds, in particular allowing for the separation of a $0\nu\beta\beta$ signal from the endpoint of the two-neutrino double beta decay ($2\nu\beta\beta$) spectrum. 

The scintillation and ionization signals are strongly anticorrelated due to large fluctuations in electron-ion recombination~\cite{Conti:2003av,Aprile:2007qd}, meaning the energy is optimally reconstructed as a linear combination of the two signals. This can be expressed as 
\begin{equation}
    E = W\cdot(S_0 + Q_0),
    \label{eq:combined_energy}
\end{equation}
where $S_0$ ($Q_0$) is the number of scintillation photons (ionization electrons) released by the event after recombination, and $W$ is a proportionality constant which represents the average energy required to produce a single quantum of either light or charge. Assuming perfect linearity and perfect anticorrelation between light and charge (that is, each electron-ion pair which recombines produces one scintillation photon), $W$ is field- and energy-independent; these assumptions are supported by measurements in Ref.~\cite{XENON:2020iwh} and Ref.~\cite{EXO-200:2019bbx}, respectively. In this picture, the energy resolution is defined by the intrinsic fluctuations in $(S_0 + Q_0)$ combined with the sources of fluctuations in the reconstruction of $S_0$ and $Q_0$ from measured quantities.

An important source of resolution broadening stems from the position-dependent detection efficiencies for both scintillation light and ionized charge. For the scintillation signals, position dependence arises from a combination of geometrical effects, surface reflectivities, and the detection efficiency of the photosensors. $S_0$ is reconstructed as
\begin{equation}
    S_0 = \frac{S_{\text{meas}}}{\epsilon_{QE}\,\times \,\epsilon_{{LM}}(x,y,z)},
    \label{eq:lightmap}
\end{equation}
where $S_\text{meas}$ is the measured number of scintillation photons, $\epsilon_{QE}$ is the photosensor quantum efficiency, and $\epsilon_{{LM}}$ is the ``lightmap'', a 3-dimensional function that describes the photon transport efficiency in the TPC.
The lightmap is a function of detector properties and may vary on timescales of several months~\cite{EXO-200:2016twj}. For the ionization signals, position dependence is driven primarily by electron attachment on electronegative impurities in the liquid xenon, which can be modeled as an exponential attenuation of the charge signal as a function of the drift time $t$. Then $Q_0$ is reconstructed as
\begin{equation}
    Q_0 = \frac{Q_{\text{meas}}}{e^{-t/\tau_e}},
    \label{eq:electron_lifetime}
\end{equation}
where $Q_\text{meas}$ is the measured number of ionization electrons, and $\tau_e$ (known as the ``electron lifetime'') represents the average time a free electron can drift in the liquid before attaching to an impurity. By design, the drift field in the TPC's active volume is uniform, meaning $t$ is proportional to the $z$ position of the event, i.e. $t = z / v_d$, where $v_d$ is the drift velocity.\footnote{Here we assume that the electric field is constant throughout the active region of the TPC. Non-uniformities in the field would create non-uniforimities in both the recombination fraction and the drift velocity $v_d$, which would introduce additional second-order position dependencies in the detected light and charge signals.}  The purity of the liquid xenon is highly correlated with operation of the recirculation system and can vary on timescales of $O(1)$~day~\cite{Ackerman:2021ijn}. Optimizing the energy reconstruction relies on the ability to measure both $\epsilon_{LM}$ and $\tau_e$ with appropriate frequency and accuracy using calibration data.  

The simplest technique for measuring these two quantities is to use a source of ionizing radiation to create events of fixed energy throughout the TPC. A standard technique is to use $\gamma$-ray sources positioned next to the detector. While this is the baseline method for nEXO, the powerful self-shielding of liquid xenon means that, to achieve sufficient statistics in the center of the TPC, the readout system needs to cope with high rates and pileup at the edges and may require long calibration campaigns.
An alternative strategy is to use radioisotopes that can be mixed directly into the active liquid xenon. Several such sources have been used previously for calibrating liquid xenon detectors, namely \isotope{Kr}{83m}~\cite{Manalaysay:2009yq,Kastens:2009pa,Akerib:2017eql}, tritiated methane~\cite{LUXTritium}, \isotope{Ar}{37}~\cite{Boulton:2017hub}, the neutron-activated isomers \isotope{Xe}{129m} and \isotope{Xe}{131m}~\cite{Ni:2007ih}, and \isotope{Rn}{220}~\cite{LangRn220,Aprile:2016pmc}. However, the first five produce signals at or below nEXO's $\sim$200~keV trigger threshold. While $^{220}$Rn is indeed under consideration for use in nEXO, a) the short half-life of the decay chain (dominated by the \SI{10.6}{hr} half-life of \isotope{Pb}{212}) is comparable to the xenon recirculation time and may affect the uniformity with which it can distribute through the TPC, and b) \isotope{Tl}{208} $\beta$-decay in the \isotope{Rn}{220} decay chain interferes with the $0\nu\beta\beta$ $Q$-value, limiting the frequency with which that source could be used. It is therefore of interest to identify other radioisotopes that can both mix throughout the TPC and produce monoenergetic signals above nEXO's trigger threshold.

\begin{figure}[!t]
    \centering
    \includegraphics[width=0.5\textwidth]{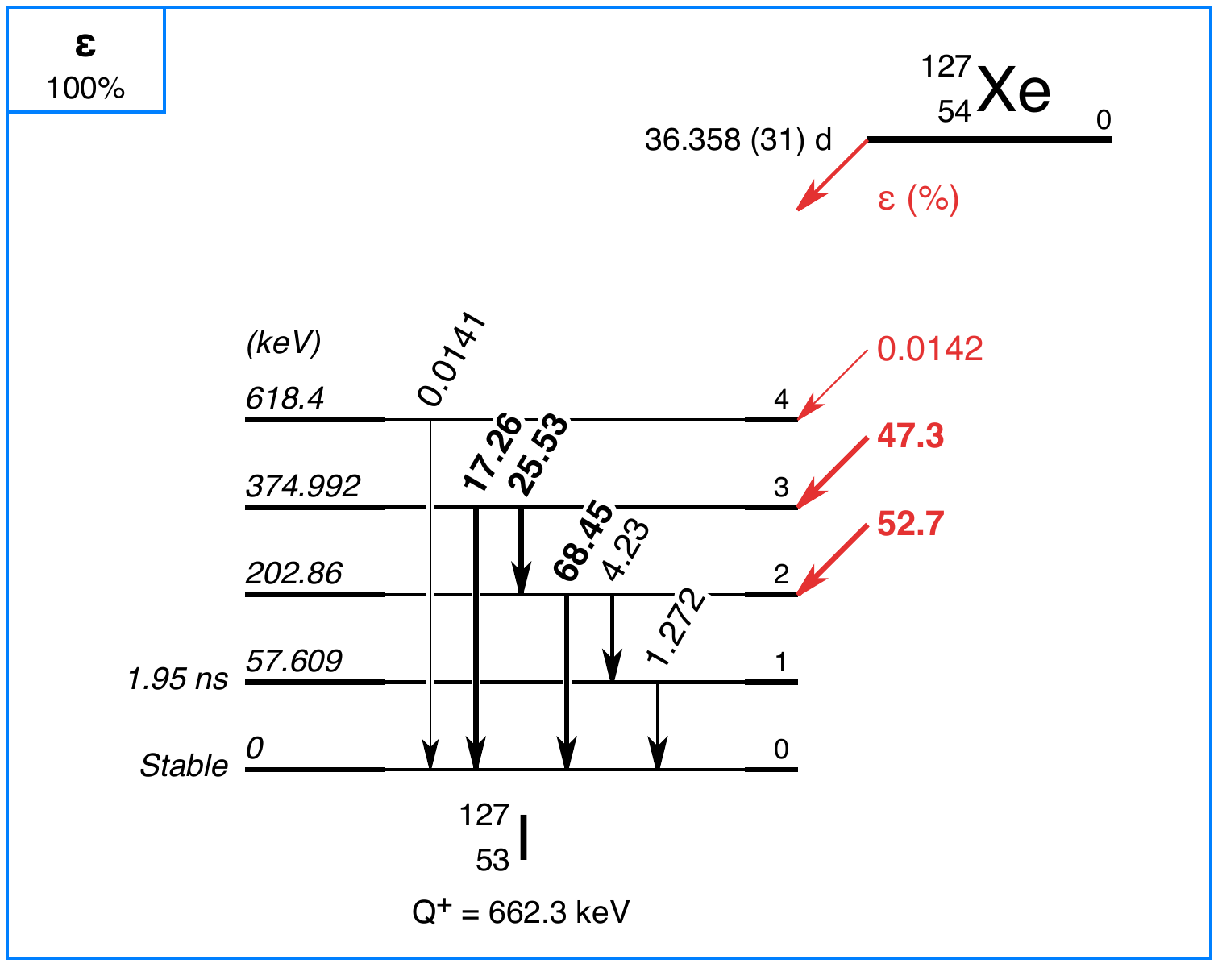}
    \caption[]{Decay scheme of \isotope{Xe}{127}, from~\cite{TabRad_v8}.\footnotemark  \isotope{Xe}{127} decays through electron capture to \isotope{I}{127}, primarily populating the levels at either $374.991(12)$~keV or $202.860(10)$~keV, which relax to the ground state via $\gamma$-ray emission. Electron capture from the K-shell and L-shell have a branching ratios of 84.2(8)\% and 12.9(1)\% and result in the emission of an additional 33.045~keV or 5.185~keV in the form of x-rays and/or Auger electrons. Most decays therefore deposit a total of either 408~keV or 236~keV in the liquid xenon, with branching fractions of 40\% and 44\%, respectively.}
    \label{fig:xe127_decay_scheme}
\end{figure}

Here we study the use of $^{127}$Xe as an injected calibration source for nEXO. This isotope decays to \isotope{I}{127} via electron capture (EC), predominantly releasing a total of either 236~keV or 408~keV of ionizing radiation that can be used for calibration. The full decay scheme is shown in Figure~\ref{fig:xe127_decay_scheme}. While this isotope has been used previously to characterize the scintillation and ionization yields of liquid xenon~\cite{LUX_Xe127_Calibration,Temples2021}, here we specifically study its use for calibrating position-dependent detection efficiencies in large-scale detectors. The 36-day half-life will ensure that the source has sufficient time to mix uniformly throughout the TPC, and the 662.3(20)~keV $Q$-value ensures that these events do not produce enough energy to interfere with the $0\nu\beta\beta$ search.
We consider a calibration strategy in which a constant activity of $\sim$1~Bq is maintained continuously in the nEXO TPC via frequent, controlled injections of \isotope{Xe}{127} into the xenon recirculation loop. This activity is similar to the expected overall background rate in nEXO (the $2\nu\beta\beta$ decay of \isotope{Xe}{136} alone produces 0.2~Bq), and will add negligible dead time. Calibration data could then be taken concurrently with physics data, providing quasi-real-time information on the detector response.

\footnotetext{Figure produced using the Laraweb tool: http://www.nucleide.org/Laraweb/index.php}

First, we discuss production of the source via neutron activation at a research reactor. Next, we demonstrate its use for measuring the electron lifetime in a prototype liquid xenon detector. Finally, we describe simulations of \isotope{Xe}{127} decays with a detailed model of the nEXO TPC and estimate the precision with which such a source could calibrate both the lightmap and the electron lifetime for different integration times.

\section{Source production}
\label{sec:production}

\subsection{Procedure for neutron activation of $^{\text{nat}}$Xe}

\begin{figure}[t]
    \centering
    \begin{subfigure}[b]{0.58\textwidth}
        \includegraphics[width=\textwidth]{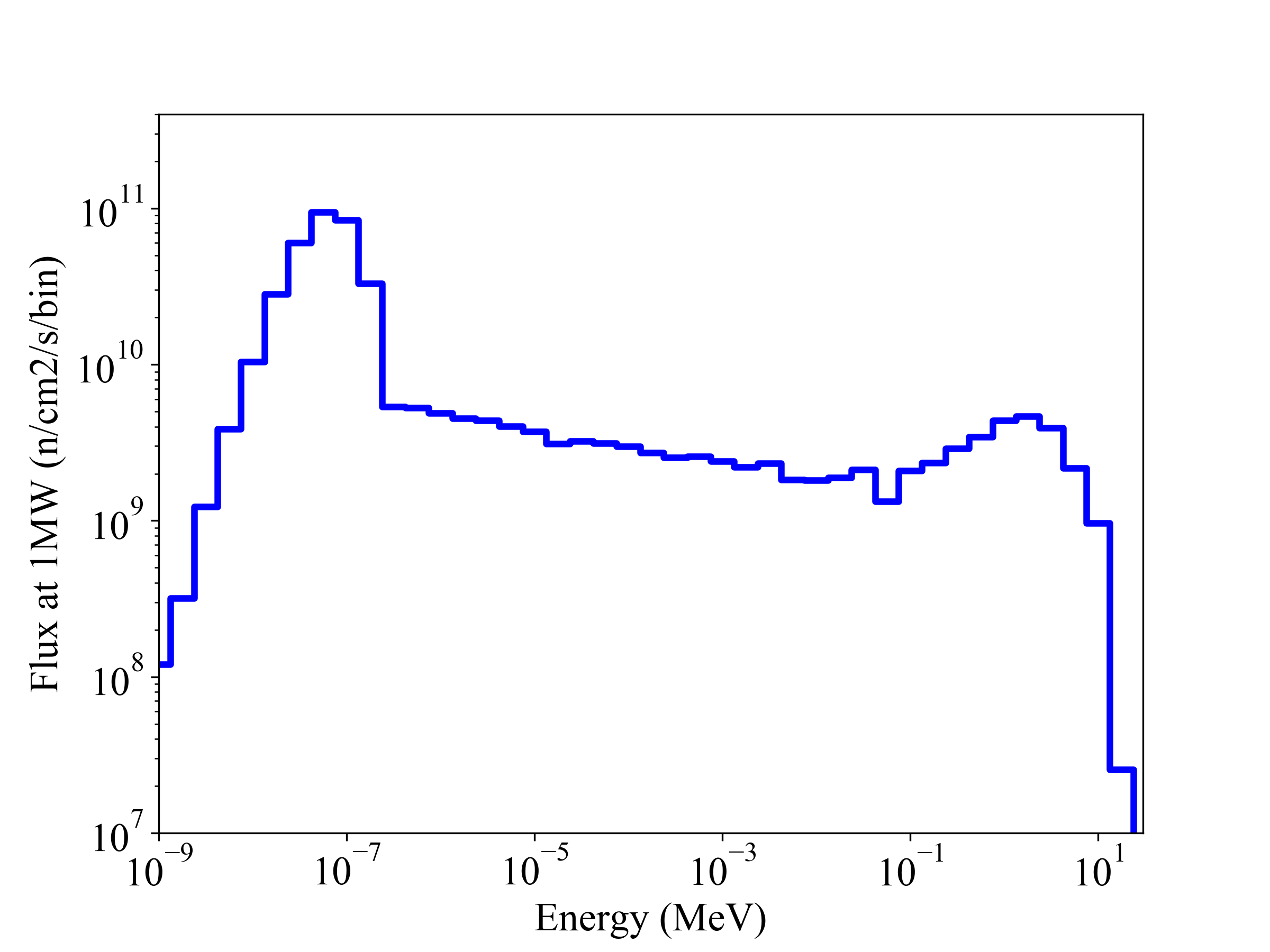}
        \caption{}\label{fig:activation_neutron_spectrum}
    \end{subfigure}
    \begin{subfigure}[b]{0.49\textwidth}
        \includegraphics[width=\textwidth]{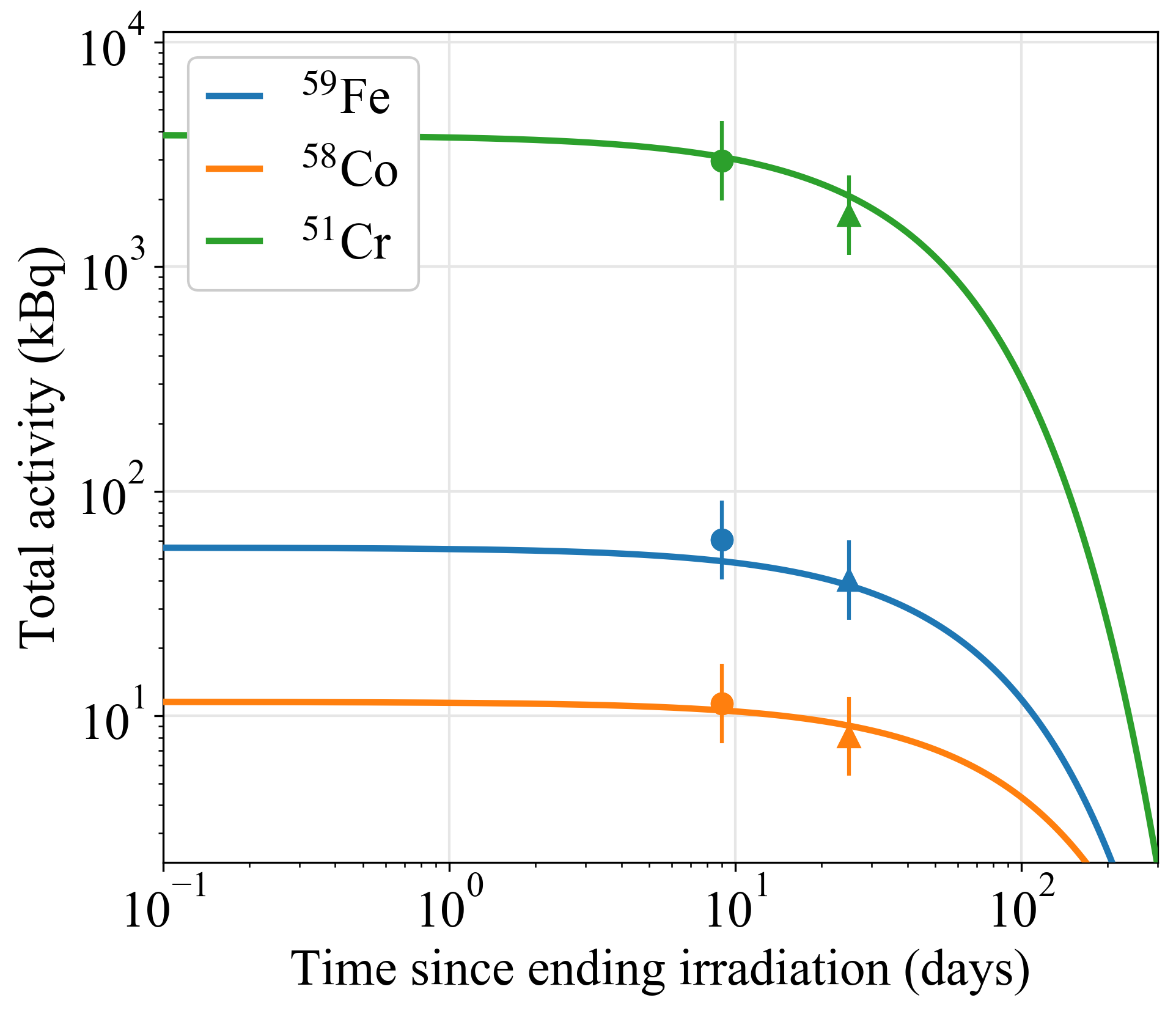}
        \caption{}\label{fig:activation_steel_activity}
    \end{subfigure}
    \begin{subfigure}[b]{0.49\textwidth}
        \includegraphics[width=\textwidth]{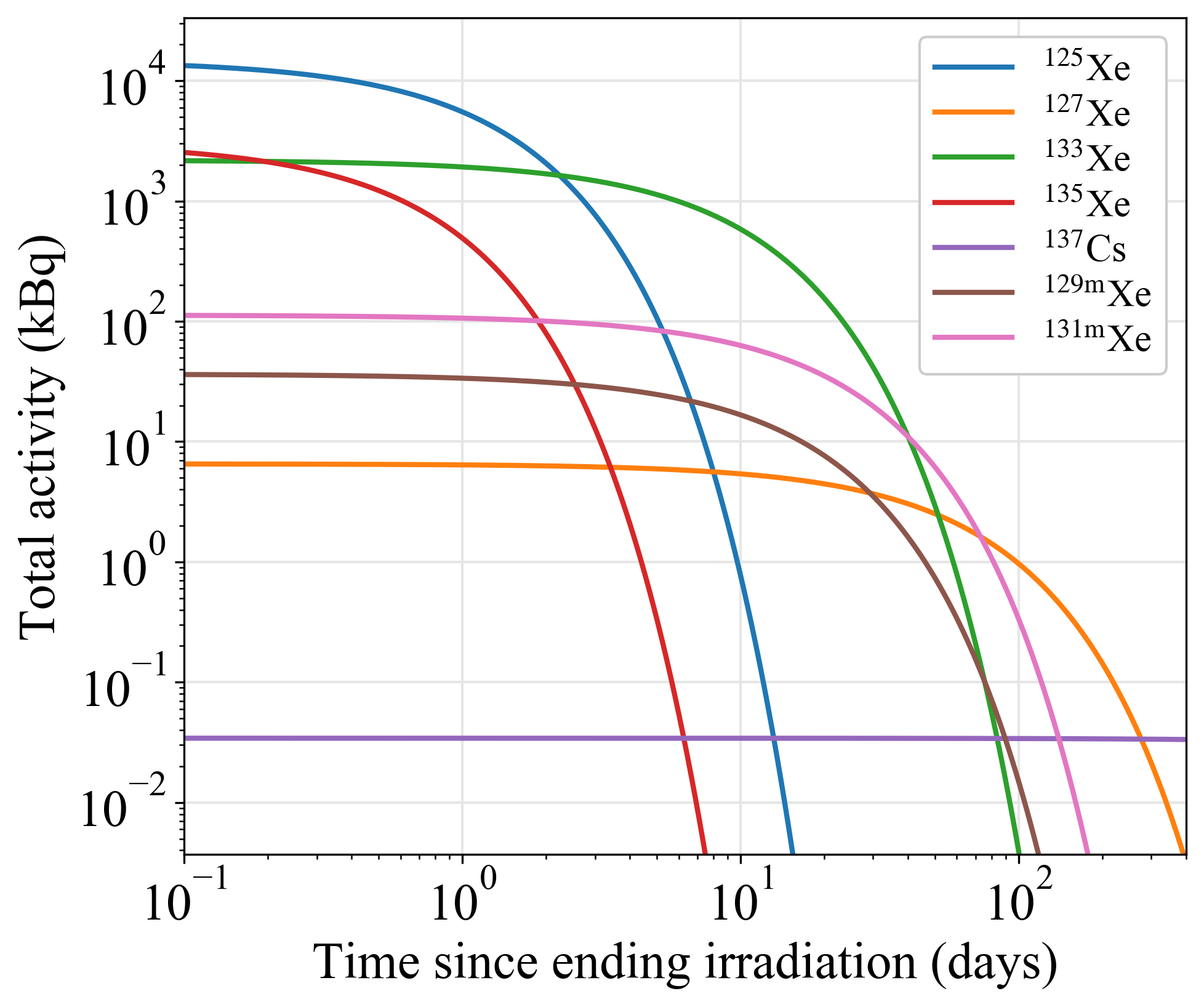}
        \caption{}\label{fig:activation_xe_activity}
    \end{subfigure}
    \caption{The expected neutron spectrum and flux at MNRC~\cite{mnrc_specs} is shown in (a), along with neutron-induced activities of various radioisotopes in both the stainless steel cylinder (b) and the xenon gas (c). The overall scaling of the neutron spectrum is calibrated using two separate measurements of the stainless steel taken at MNRC (b, circles) and Stanford University (b, triangles), as explained in the text. The error bars are dominated by the estimated 50\% systematic uncertainty in each measurement. This measured neutron flux is used to calculate the expected activity vs. time (solid lines). }
    \label{fig:activation}
\end{figure}

\isotope{Xe}{127} can be readily produced via neutron capture on \isotope{Xe}{126}, which is present in natural xenon at an isotopic abundance of $\sim$0.1\%. 
To produce the source, a 150~cm$^3$ stainless steel (316L) sample cylinder was filled with \SI{69}{g} of \isotope{Xe}{nat} gas and shipped to the research reactor at McClellan Nuclear Research Center (MNRC). The sample cylinder was placed in the Neutron Transmutation Doping (NTD) void, for which the steady-state neutron spectrum at 1~MW is given in Ref~\cite{mnrc_specs} and shown in Figure~\ref{fig:activation_neutron_spectrum}. 
The irradiation was performed for 15 minutes at a power of $\sim$\SI{250}{kW}. The production of radioisotopes in both the stainless steel cylinder and the \isotope{Xe}{nat} was calculated numerically by folding the expected reactor spectrum with cross sections from standard libraries. Neutron capture cross sections for most isotopes were taken from \uppercase{ENDF/B-VII.0}~\cite{ENDF7}. For the two metastable isomers, \isotope{Xe}{129m} and \isotope{Xe}{131m}, we used cross sections from the TENDL-2019 library~\cite{KONING20191}, which are conveniently given in terms of the total cross section for populating the desired state given a specific reaction.

The neutron flux incident on the sample was calibrated using activation in the stainless steel cylinder, which produces three long-lived products: \isotope{Cr}{51} and \isotope{Fe}{58}, which are produced primarily by thermal neutron capture, and \isotope{Co}{58}, which is produced primarily by fast neutron $(n,p)$ reactions. Two short ($\sim$30~min) radioassay measurements of the cylinder using high-purity germanium (HPGe) counters were performed: the first was taken at MNRC 9~days after irradiation; the second was taken at Stanford University 25~days after irradiation. We estimate $\sim$50\% systematic uncertainties in each measurement due to uncertainties in the counting geometry. The measured activities are listed in Table~\ref{tab:stainless_steel_activation} and plotted in Figure~\ref{fig:activation_steel_activity}. We infer the overall neutron flux by scaling the neutron spectrum in the calculations described above so that the predicted activities match the measured activities.
While the estimated fluxes from the measurements at 25~days are systematically lower than those at 9~days by approximately 30\%, both measurements are consistent within the estimated uncertainties.
Importantly, the neutron fluxes inferred from the thermal-neutron-induced reactions are consistent with those inferred from the fast neutron reactions, indicating that the assumed spectral shape is adequate. 
Taking the average and standard deviation of all inferred values gives an estimated neutron flux of $(2.1 \pm 0.3)\times10^{10}$~n/cm$^2$/s. We note that, for unclear reasons, this flux is lower by a factor of $\sim$5 than the expected steady-state flux at a reactor power of \SI{250}{kW}. We attribute this discrepancy to operating the reactor in a transient mode during irradiation of our sample.

The predicted activities of radioisotopes produced in the \isotope{Xe}{nat} gas, given the flux estimated above, are shown in Figure~\ref{fig:activation_xe_activity}. The activity in the \isotope{Xe}{nat} gas sample shortly after irradiation is dominated by short-lived isotopes. Of these, the longest-lived are the metastable isomers \isotope{Xe}{129m} and \isotope{Xe}{131m}. Thermal neutron capture is expected to be the dominant production mechanism for these isotopes at a reactor, in contrast with the \isotope{Cf}{252}-based activation scheme reported in Ref.~\cite{Ni:2007ih}. 
Despite the high initial activity, after a cool-off period of $\sim$100~days the remaining activity is dominated by \isotope{Xe}{127}. We note that we predict a non-negligible amount of long-lived \isotope{Cs}{137} produced by neutron capture on \isotope{Xe}{136} followed by beta decay of \isotope{Xe}{137} ($T_{1/2}$ = 3.8~min), but Cs is expected to be easily removed from the Xe gas by standard purification techniques prior to a deployment in nEXO. 

\begin{table}[t]
    \footnotesize

    \renewcommand{\arraystretch}{1.3}

    \caption{Radioisotopes produced by neutron activation of stainless steel, which we use to calibrate the neutron flux incident on the sample. The value of $\Delta T$ denotes the elapsed time between irradiation and when the measurements were made. Systematic uncertainties in the measurements are estimated to be $\pm$50\%, due to uncertainties in the counting geometry. Statistical uncertainties are $O(1)$\% and are omitted for clarity.}
    \label{tab:stainless_steel_activation}
    
    \centering
    
    \begin{tabular}{c|c|c|c|c|c|c}
        \hline
        Isotope & Half-life &Production mode & $\gamma$-ray energy & $\Delta T$ & Meas. activity & Inferred flux  \\
        \hline
        \hline
        \multirow{2}{*}{\isotope{Cr}{51}} & \multirow{2}{*}{27.7~d} & \multirow{2}{*}{ \isotope{Cr}{50} $(n,\gamma)$ \isotope{Cr}{51} } & \multirow{2}{*}{320~keV} & 9~d &  2960~kBq & $2.3 \times 10^{10}$~n/cm$^2$/s \\ 
         & & & & 25~d & 1690~kBq &  $1.7 \times 10^{10}$~n/cm$^2$/s \\ 
        \hline
        \multirow{2}{*}{\isotope{Fe}{59}} & \multirow{2}{*}{44.5~d} & \multirow{2}{*}{\isotope{Fe}{58} $(n,\gamma)$ \isotope{Fe}{59}} & \multirow{2}{*}{1099 \& 1291~keV} & 9~d &  61~kBq & $2.7 \times 10^{10}$~n/cm$^2$/s \\
         & & & & 25~d & 40~kBq & $2.3 \times 10^{10}$~n/cm$^2$/s\\
        \hline
        \multirow{2}{*}{\isotope{Co}{58}} & \multirow{2}{*}{70.9~d} & \multirow{2}{*}{\isotope{Ni}{58} $(n,p)$ \isotope{Co}{58}} & \multirow{2}{*}{810~keV} & 9~d & 11.3~kBq & $2.3 \times 10^{10}$~n/cm$^2$/s \\
         & & & & 25~d & 8.1~kBq & $1.8 \times 10^{10}$~n/cm$^2$/s \\
        \hline
    \end{tabular}

\end{table}

\subsection{Low-background radioassay measurements of activated gas}
\label{subsec:low_bkg_radioassay}

\begin{figure}
    \begin{subfigure}{0.49\textwidth}
    \centering
    \includegraphics[width=0.7\textwidth]{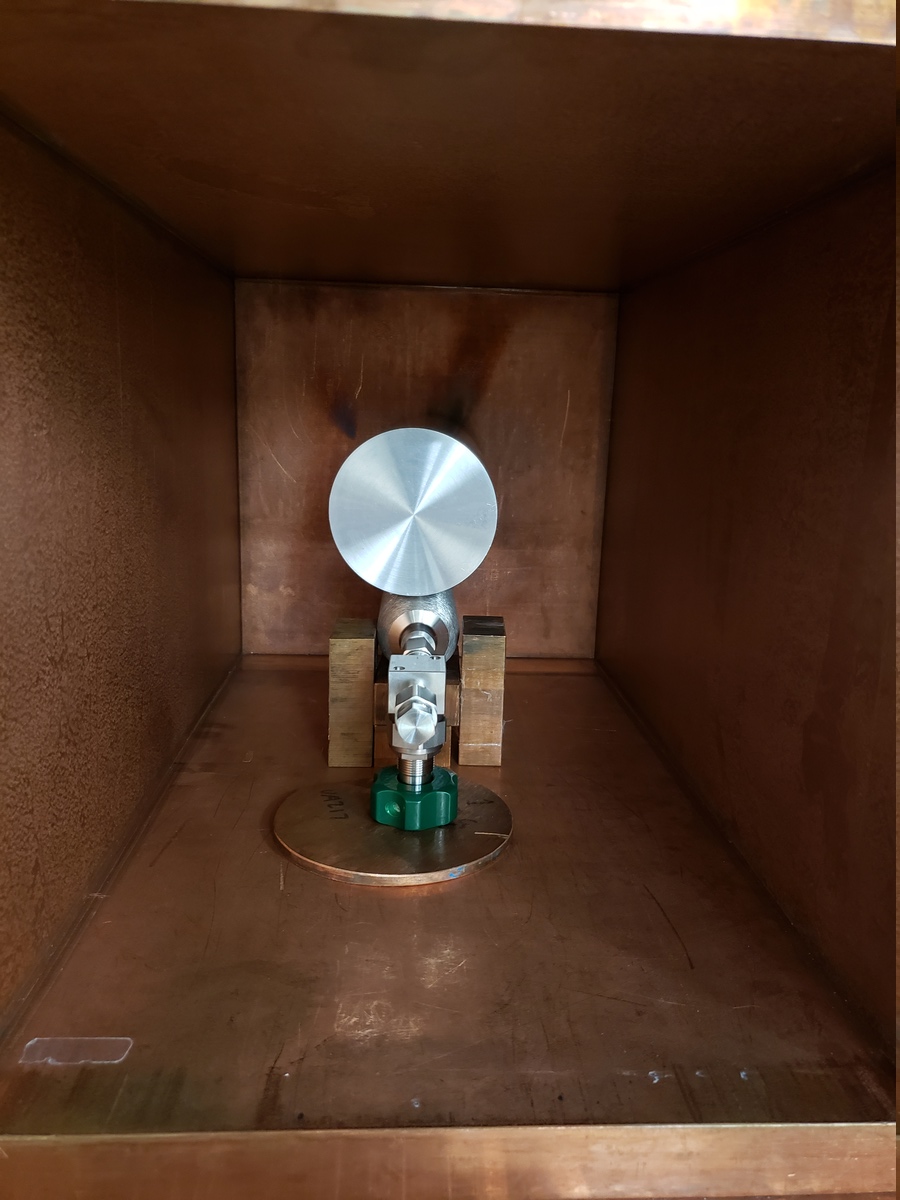}  
    \caption{Counting setup}
    \label{fig:ua_counting_ge3}
    \end{subfigure}
    \begin{subfigure}{0.49\textwidth}
    \centering
    \includegraphics[width=\textwidth]{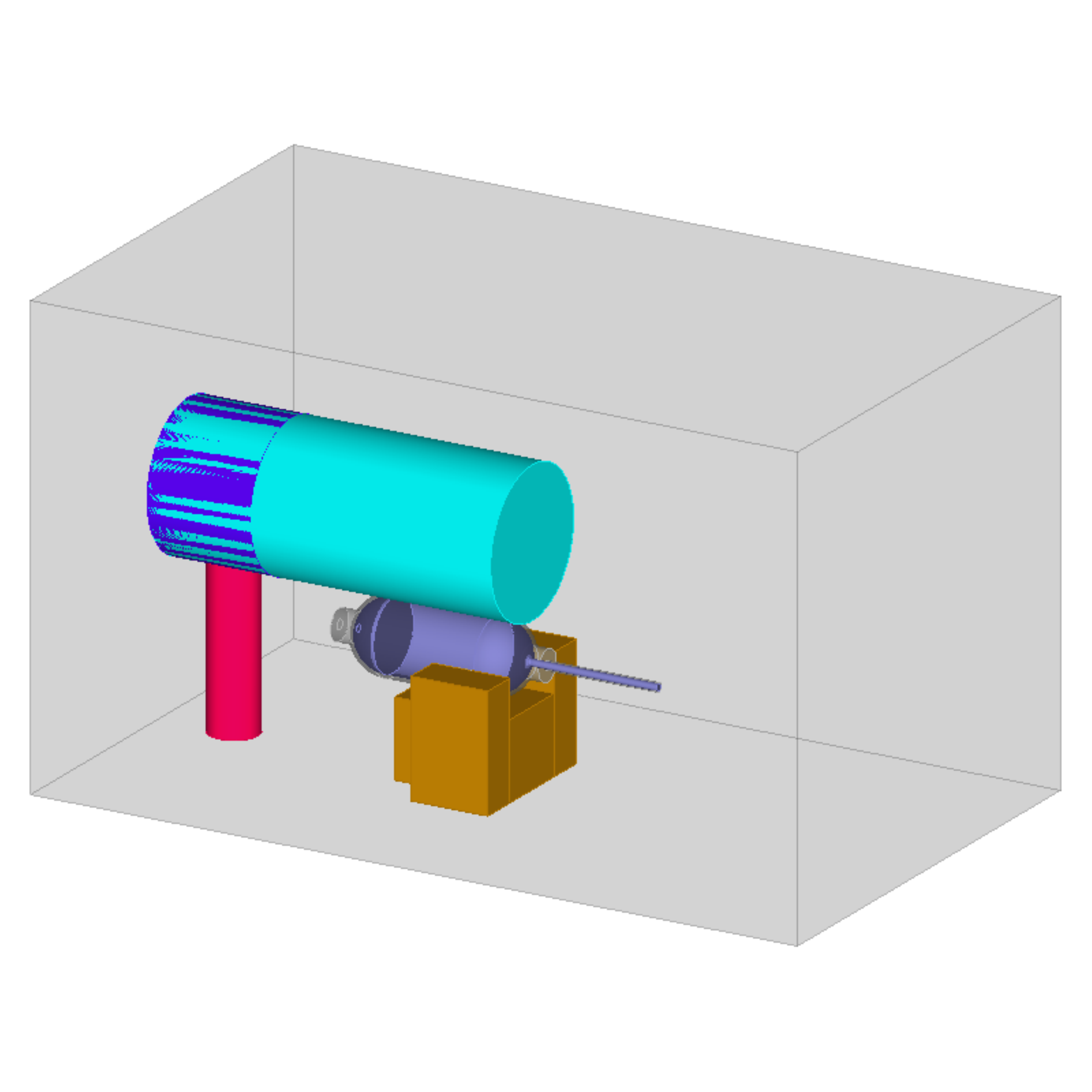} 
    \caption{GEANT4 rendering}
    \label{fig:ua_counting_g4}
    \end{subfigure}
    \caption{Activated Xe gas being counted at the University of Alabama. The cylinder was placed directly below the HPGe detector, inside low-background copper shielding. The GEANT4 model shown in (b) is used to calculate the $\gamma$-ray detection efficiency of the counting geometry.}
    \label{fig:ua_counting_setup}
\end{figure}

\begin{figure}
    \centering
    \includegraphics[width=0.9\textwidth]{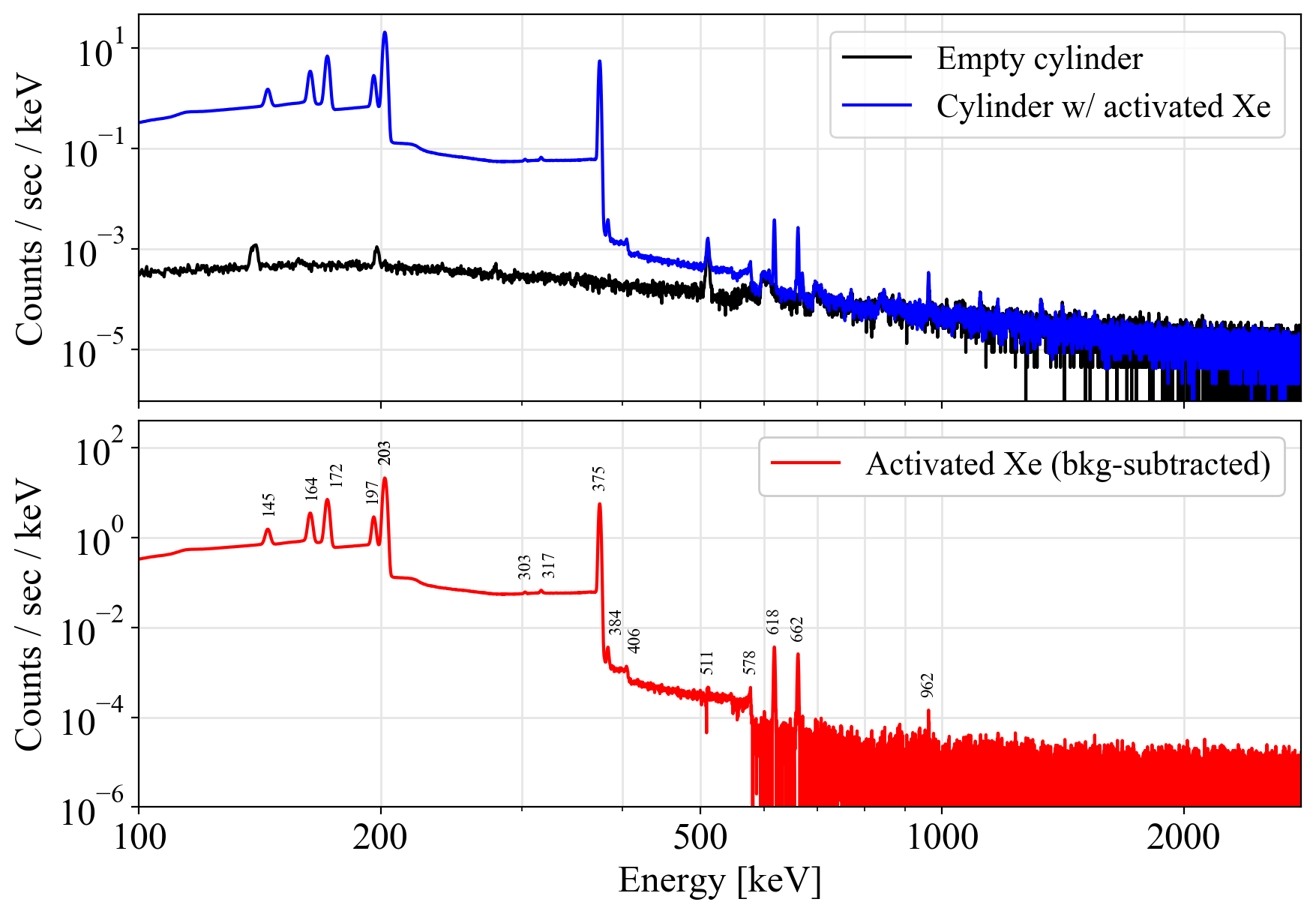}
    \caption{$\gamma$-ray energy spectrum of the activated Xe gas measured in a low-background HPGe detector at the University of Alabama. The raw spectra (top) consist of two measurements of an un-activated stainless steel sample cylinder, the first under vacuum (black) and the second filled with the activated Xe gas (blue). The former is subtracted from the latter to get the $\gamma$-ray spectrum of the activated gas sample (bottom, red). }
    \label{fig:radioassay_energy_spectra}
\end{figure}

High-precision radioassay measurements of the activated Xe gas were made by transferring the gas into an unactivated cylinder and counting the sample with a low-background HPGe detector at the University of Alabama~\cite{Tsang:2019apx}. A photo of the experimental setup is shown in Figure~\ref{fig:ua_counting_ge3}.
The purpose of these measurements was twofold: first, we compare measured and predicted activities in the gas to estimate the accuracy of our calculations, which requires more sensitive measurements than those of the steel due to the smaller amount of material; second, we use these data to search for any unexpected activation products in the gas that could produce unwanted backgrounds in nEXO. Before filling, the new cylinder was counted for approximately two weeks to obtain a background spectrum. The cylinder was then filled and counted for two more weeks. In each case, data were acquired in 4-hour intervals.
The total energy spectra, summed across the entire campaign, are shown in Figure~\ref{fig:radioassay_energy_spectra}. 

We used a moving average window technique to detect peaks in the summed spectrum, then fit them in each of the 4-hour time slices using a Gaussian line shape plus a linear background. 
The fitted peak areas as a function of time were then fitted to an exponential model to extract the mean lifetime of the nuclides. The fitted energy and mean lifetime were compared to the Evaluated Nuclear Structure Data Files (ENSDF, Ref.~\cite{nndc}) to identify a candidate nuclide, then an activity of the nuclide in the $i$th time slice, $A_i$, was determined by the following formula:
\begin{equation}
    A_i = \frac{C_i}{\varepsilon \cdot b \cdot \Delta t}
\end{equation}
where $C_i$ is the number of counts registered, $\Delta t =  4$ hours, $b$ is the tabulated $\gamma$-ray intensity from NNDC, and $\varepsilon$ is the energy-dependent $\gamma$-ray detection efficiency. The efficiency estimation, which includes modeling of the detector dead layer, was performed with calibration sources and the Geant4-based GeSim package~\cite{Tsang:2019apx} (the GeSim rendering of the counting setup is shown in Figure \ref{fig:ua_counting_g4}). For point sources, the systematic uncertainty in $\varepsilon$ is 9\%. The dead-time and pileup in the HPGe detector were negligible in these measurements. 
The measured activities are shown in Table~\ref{tab:xe_activation}.

\begin{table}
\footnotesize
\renewcommand{\arraystretch}{1.2}
\caption{Measurements of long-lived ($T_{1/2} > 1$~d) radioisotopes produced by neutron activation of \isotope{Xe}{nat}. Activities are reported in terms of the activity immediately following the irradiation campaign, as explained in the text. The predicted activity uses the calibrated neutron flux obtained from measurements of the stainless steel (Table~\ref{tab:stainless_steel_activation}). Statistical uncertainties on the measured activities are $O(0.1)$\% and are omitted for clarity. Systematic uncertainties in the measurements are $O(10)$\% and are discussed in the text (Section~\ref{subsec:low_bkg_radioassay}).}
\label{tab:xe_activation}
\centering
\begin{threeparttable}
\begin{tabular}{c|c|c|c|c|c}
    
    \hline
    Isotope & Half-life & Production mode & $\gamma$-ray energy & Measured activity & Predicted activity  \\
    \hline
    \hline
    \multirow{8}{*}{\isotope{Xe}{127}} & \multirow{8}{*}{36.3~d} & \multirow{8}{*}{ \isotope{Xe}{126} $(n,\gamma)$ \isotope{Xe}{127} } & 145~keV & 4.8~kBq & \multirow{8}{*}{6.5~kBq}  \\
     & & & 172~keV & 5.0~kBq &  \\
     & & & 203~keV & 5.5~kBq &  \\
     & & & 317~keV\tnote{*} & - &  \\
     & & & 375~keV & 6.3~kBq &  \\
     & & & 406~keV\tnote{*} & - &  \\
     & & & 578~keV\tnote{*} & - &  \\
     & & & 618~keV & 6.9~kBq &  \\
     \hline
    \multirow{2}{*}{\isotope{Xe}{129\text{m}}} & \multirow{2}{*}{8.88~d} & \isotope{Xe}{128} $(n,\gamma)$ \isotope{Xe}{129\text{m}} & \multirow{2}{*}{197~keV} & \multirow{2}{*}{137~kBq}& \multirow{2}{*}{36~kBq} \\
    & & \isotope{Xe}{129} $(n,n')$ \isotope{Xe}{129\text{m}} & & \\
    \hline
    \multirow{2}{*}{\isotope{Xe}{131\text{m}}} & \multirow{2}{*}{11.8~d} & \isotope{Xe}{130} $(n,\gamma)$ \isotope{Xe}{131\text{m}} & \multirow{2}{*}{164~keV} & \multirow{2}{*}{187~kBq} & \multirow{2}{*}{113~kBq} \\
    & & \isotope{Xe}{131} $(n,n')$ \isotope{Xe}{131\text{m}} & & \\
    \hline
    \multirow{3}{*}{\isotope{Xe}{133}} & \multirow{3}{*}{5.25~d} &  \multirow{2}{*}{\isotope{Xe}{132} $(n,\gamma)$ \isotope{Xe}{133}} &  303~keV & 2170~kBq & \multirow{2}{*}{2190~kBq} \\
    & & & 384~keV & 2250~kBq &  \\
    \hline
    \multirow{2}{*}{\isotope{Cs}{137}} & \multirow{2}{*}{30.1~y} & \isotope{Xe}{136} $(n,\gamma)$ \isotope{Xe}{137}, &  \multirow{2}{*}{662~keV} & \multirow{2}{*}{ $3.1\times10^{-4}$~kBq} & \multirow{2}{*}{$3.4\times10^{-2}$~kBq} \\
    & & \isotope{Xe}{137} $\rightarrow$ \isotope{Cs}{137} & & & \\
     \hline
\end{tabular}
\begin{tablenotes}
    \item[*] Features at these energies are produced by pile-up of lower-energy $\gamma$-rays.
\end{tablenotes}
\end{threeparttable}
\end{table}

The measured activities are generally in agreement with the predictions, based on the neutron flux evaluation described previously. For \isotope{Xe}{127}, the measured activity using different emission lines spans 4.8~--~6.9~kBq, in agreement with the predicted value at the level of $\sim$30\%. The measured values are systematically lower than the predictions when the gamma energy goes below $\sim$200~keV; we attribute this to additional systematic error in the estimated $\gamma$-ray detection efficiency (beyond the expected 9\%) due to simplifications in the simulated geometry of the gas cylinder, which will affect efficiency estimates for lower-energy $\gamma$-rays more strongly due to their higher attenuation in materials. The activities derived from the lines at 375 and 618~keV, for which the efficiencies are easier to estimate, agree with the predicted value within 5\%. Similar agreement is observed for \isotope{Xe}{133}. Larger discrepancies are observed for the metastable isomers \isotope{Xe}{129m} and \isotope{Xe}{131m}, which we attribute to uncertainties in the evaluated reaction cross sections for populating specific excited states; evaluations are extrapolated from a single measurement of \isotope{Xe}{129m} and \isotope{Xe}{131m} activation by thermal neutrons which carries 30--40\% uncertainties~\cite{KONDAIAH1968329}.
The \isotope{Cs}{137} activity exhibits the largest discrepancy, with the measured activity a factor of $\sim$100 smaller than the predicted value. We hypothesize that most of the Cs attaches to the inner surface of the activated cylinder and is consequently not transferred to the new cylinder used for counting. We take this as an encouraging sign that, using dedicated purification techniques, the \isotope{Cs}{137} contamination of such a source could be reduced to negligible levels. We find no evidence for the production of unexpected isotopes in the radioassay data. 
While we note that additional measurements may be required to ensure that such a source meets the stringent ultra-low-background requirements for nEXO, these results demonstrate that neutron activation of \isotope{Xe}{nat} gas is a promising path to producing a \isotope{Xe}{127} calibration source. 

\section{Experimental demonstration}
\label{sec:demonstration}

\subsection{Stanford LXe TPC}

The activated xenon was injected into the Stanford liquid xenon TPC to demonstrate its use as a calibration source. 

\begin{figure}

    \centering
    \begin{subfigure}[b]{0.55\textwidth}
        \centering
        \includegraphics[width=0.95\textwidth]{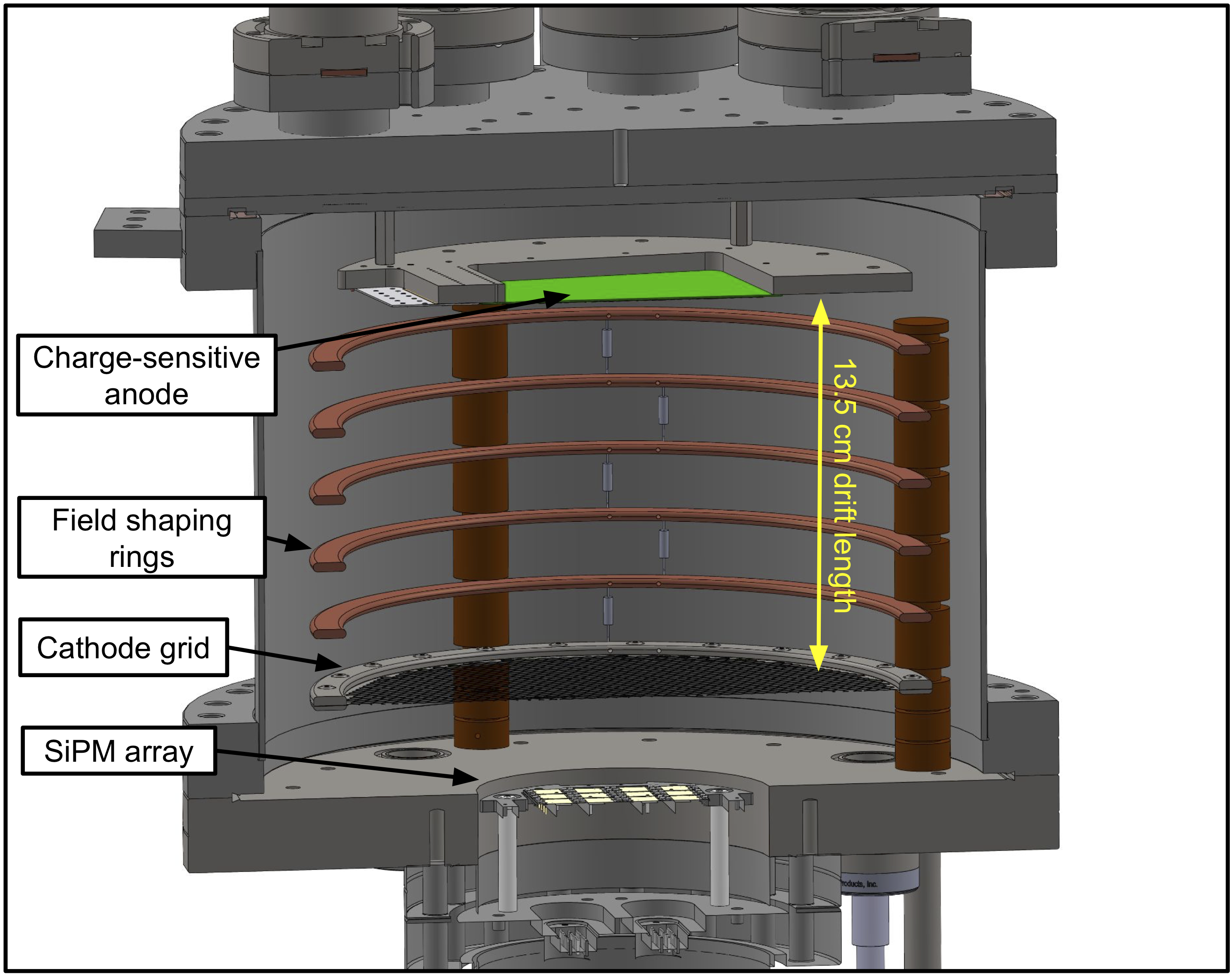}
        \caption{}\label{fig:chamber_cad:chamber}
    \end{subfigure} %
    \begin{subfigure}[b]{0.44\textwidth}
        \centering
        \includegraphics[width=0.99\textwidth]{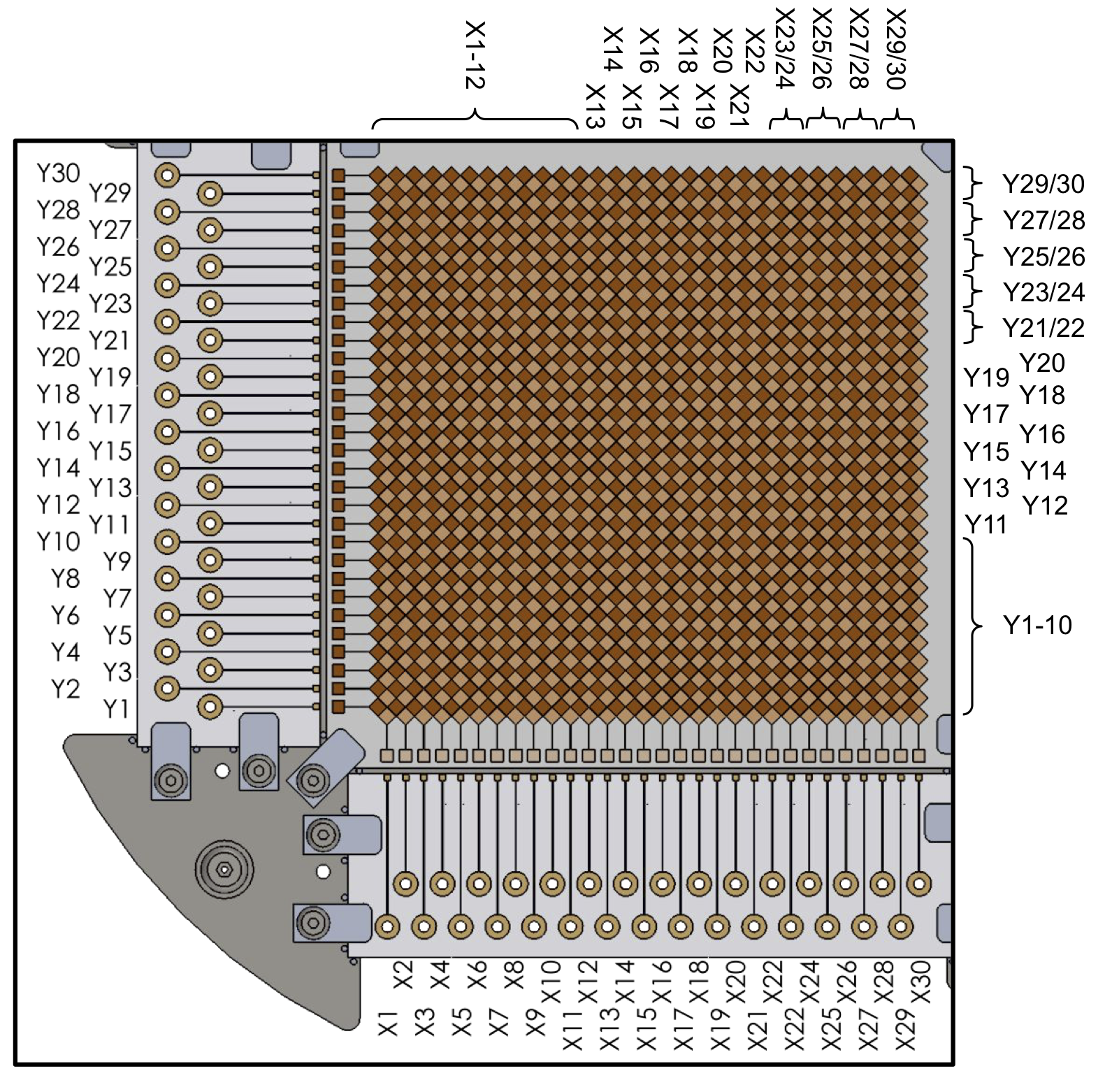}
        \caption{}\label{fig:chamber_cad:tile}
    \end{subfigure}

    \caption{(a) CAD model of chamber and tile readout. The individual, \SI{3}{mm}-pitch strips on the charge-sensing anode are ganged into channels as illustrated in (b).}
    \label{fig:chamber_cad}
\end{figure}

The TPC is housed in a cylindrical stainless steel chamber, \SI{20.3}{cm} long by \SI{25.4}{cm} diameter, maintained at \SI{165}{K} and filled with \SI{27}{kg} of liquid xenon.
The TPC itself, illustrated in Figure~\ref{fig:chamber_cad:chamber}, consists of a \SI{13.5}{cm} drift volume defined at the top by the charge-sensing anode plane and at the bottom by a stainless steel cathode grid. A uniform electric field is maintained by five field shaping rings connected by \SI{1}{\giga\ohm} resistors. During the measurements discussed here, the cathode was maintained at \SI{6}{kV}, producing an electric field of $\sim$\SI{400}{V/cm} in the active volume of the TPC. Scintillation light is detected by an array of 24 VUV-sensitive FBK VUV-HD1 silicon photomultipliers (SiPMs) which are paired into 12 readout channels, located at the bottom of the chamber approximately \SI{4}{cm} below the cathode. Ionization is detected by a prototype nEXO charge tile, described in detail in Ref.~\cite{Jewell:2017dzi}. The tile consists of square gold pads deposited on a quartz substrate, connected into strips of \SI{9}{cm} length and \SI{3}{mm} pitch in both the $x$ and $y$ dimensions. The strips are connected via feedthroughs to discrete preamplifiers (based on the design in Ref.~\cite{Fabris_1999}), operating at $\sim$\SI{165}{K} but located outside the xenon space. Due to a limited number of feedthroughs in the xenon chamber, some strips are ganged together into a single channel, as illustrated in Figure~\ref{fig:chamber_cad:tile}.  
Both the light and the charge signals are digitized by Struck SIS3316 digitizers at a rate of \SI{62.5}{MHz}. 
Data acquisition is triggered by a two-fold coincidence requirement on the SiPM channels, with the thresholds on each channel set at the mean pulse height of single photoelectrons.

\subsection{Injection procedure}

\begin{figure}
    \centering
    \begin{subfigure}[t]{0.46\linewidth}
        \centering
        \includegraphics[width=0.85\linewidth]{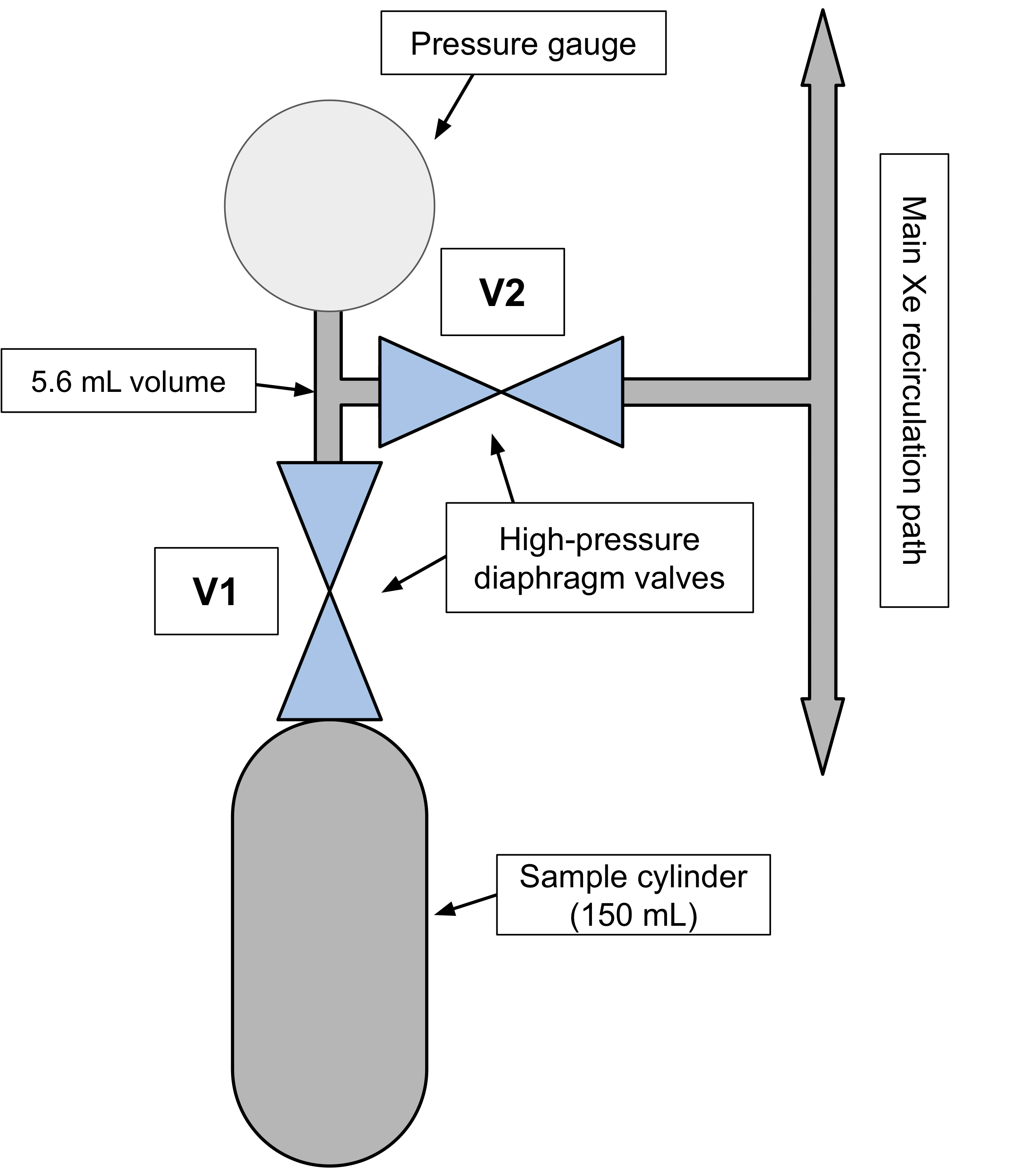}
        \caption{}
        \label{fig:injection_schematic}
    \end{subfigure}%
    \begin{subfigure}[t]{0.48\linewidth}
        \centering
        \includegraphics[width=0.95\linewidth]{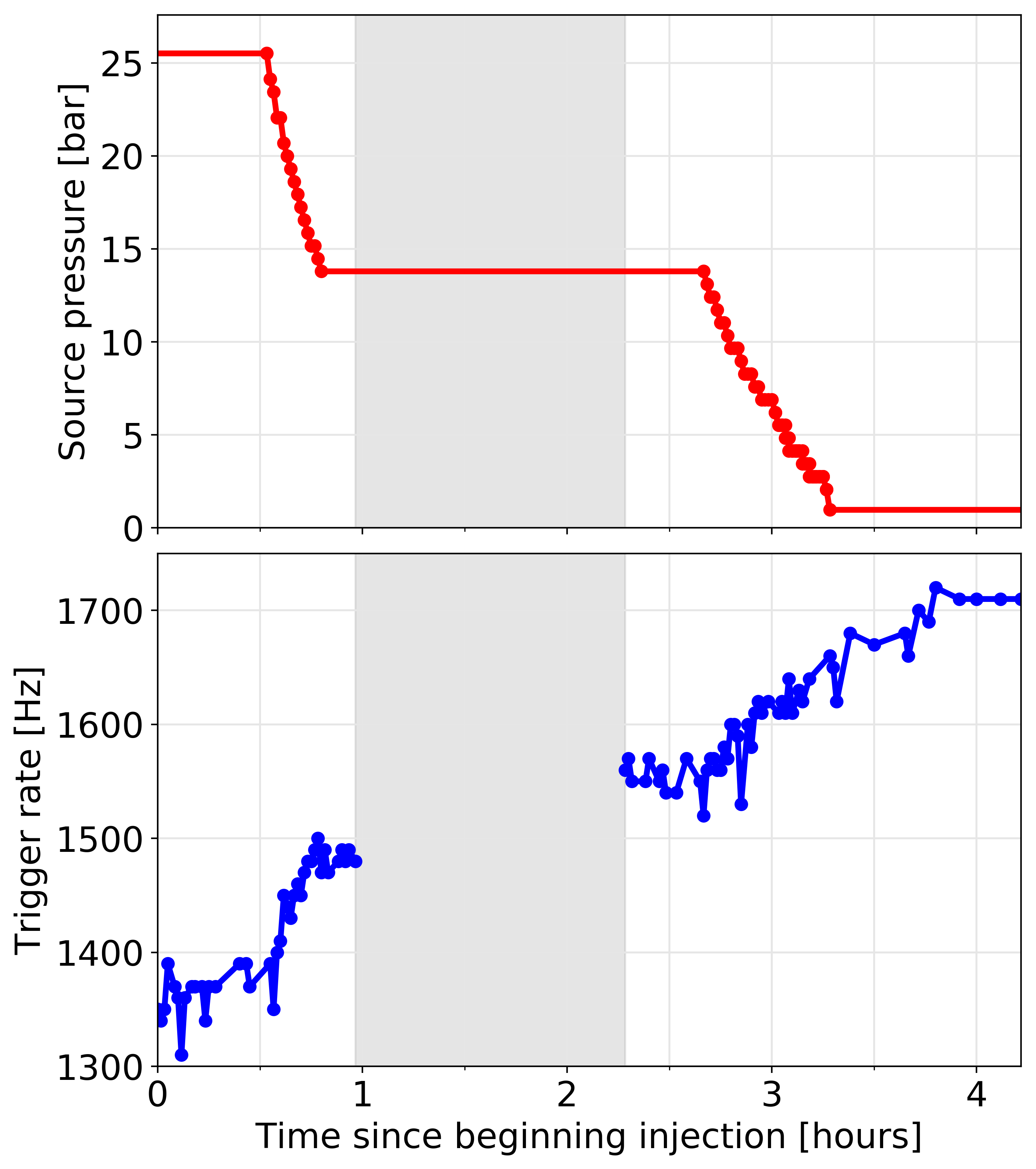}
        \caption{}
        \label{fig:injection}
    \end{subfigure}%
    
    \caption{(a) Diagram of the source injection hardware. (b) Source pressure, measured inside the 5.6~mL buffer volume during each fill/release cycle, and detector trigger rate during the $^{127}$Xe injection test campaign. The shaded grey region indicates a pause in data taking.}
    
\end{figure}

The activated xenon injection hardware is illustrated in Figure~\ref{fig:injection_schematic}. The cylinder filled with gas is connected through a valve (V1) to a tee, which connects both to a high-pressure gauge and to the main xenon recirculation loop through a second valve (V2). The 5.6~mL volume between the two valves serves as a buffer volume which can be pressurized with activated gas from the cylinder, then opened to the main xenon recirculation loop to inject the gas into the system. During each fill/release cycle we measure the pressure in the buffer volume to calculate the amount of gas injected. 

The injection test was done in two stages, each of which consisted of $\sim$20 fill/release cycles. Xenon was continuously recirculated at approximately \SI{5}{SLPM} to promote mixing.  During each injection, the SiPM trigger rate in the TPC was monitored to ensure that the activity was reaching the detector. The results are shown in Figure~\ref{fig:injection}. A clear correlation between injected gas and trigger rate is observed, indicating that at least some of the the activated gas mixed quickly into the TPC. There is also evidence of increasing activity after the injection was stopped, suggesting that the distribution of $^{127}$Xe throughout the TPC was not immediate, and that mixing continued for some time.

\subsection{Electron lifetime measurement}

After injection, the xenon was recirculated for several days to ensure that the \isotope{Xe}{127} was distributed uniformly. Recirculation was then stopped and data were taken in four separate acquisitions over the course of two days to measure the electron lifetime in the TPC by mesuring the detected ionization signal as a function of the drift time. 

Charge collection signals on individual channels are included in our analysis if they are above 3 times the RMS of the baseline noise. The charge drift time is computed as the time difference between the 90\% rise time of the waveform and the scintillation trigger. Signals are first grouped into ``clusters'' if their reconstructed drift times are within 3~$\mu$s of each other. We then apply an event selection cut that requires each cluster to be reconstructed on at least one $x$ and $y$ channel to ensure that events are fully reconstructed in all three spatial dimensions. Finally, we select events for which the charge-weighted average position in the $x$-$y$ plane falls in the central $\pm$15~mm of the TPC, to select the region of the TPC with the finest-grained position resolution. The total ionization energy of the event is reconstructed by summing the charge from all the detected signals. 

\begin{figure}[t]
    \centering
    \begin{subfigure}[t]{0.47\linewidth}
        \centering
        \includegraphics[width=\textwidth]{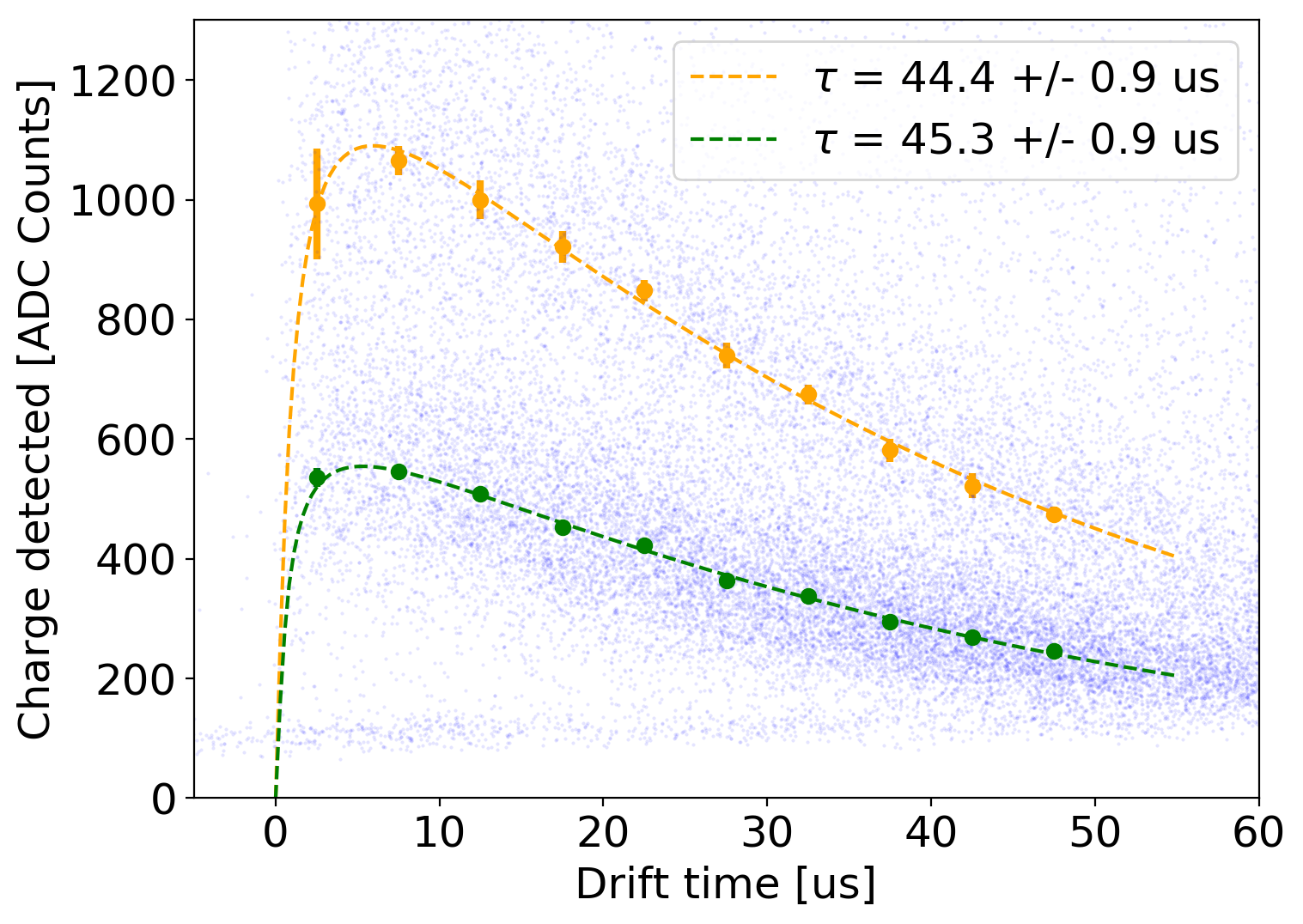}
        \caption{}
        \label{fig:charge_vs_drift_time}
    \end{subfigure}
    \begin{subfigure}[t]{0.5\linewidth}
        \includegraphics[width=\textwidth]{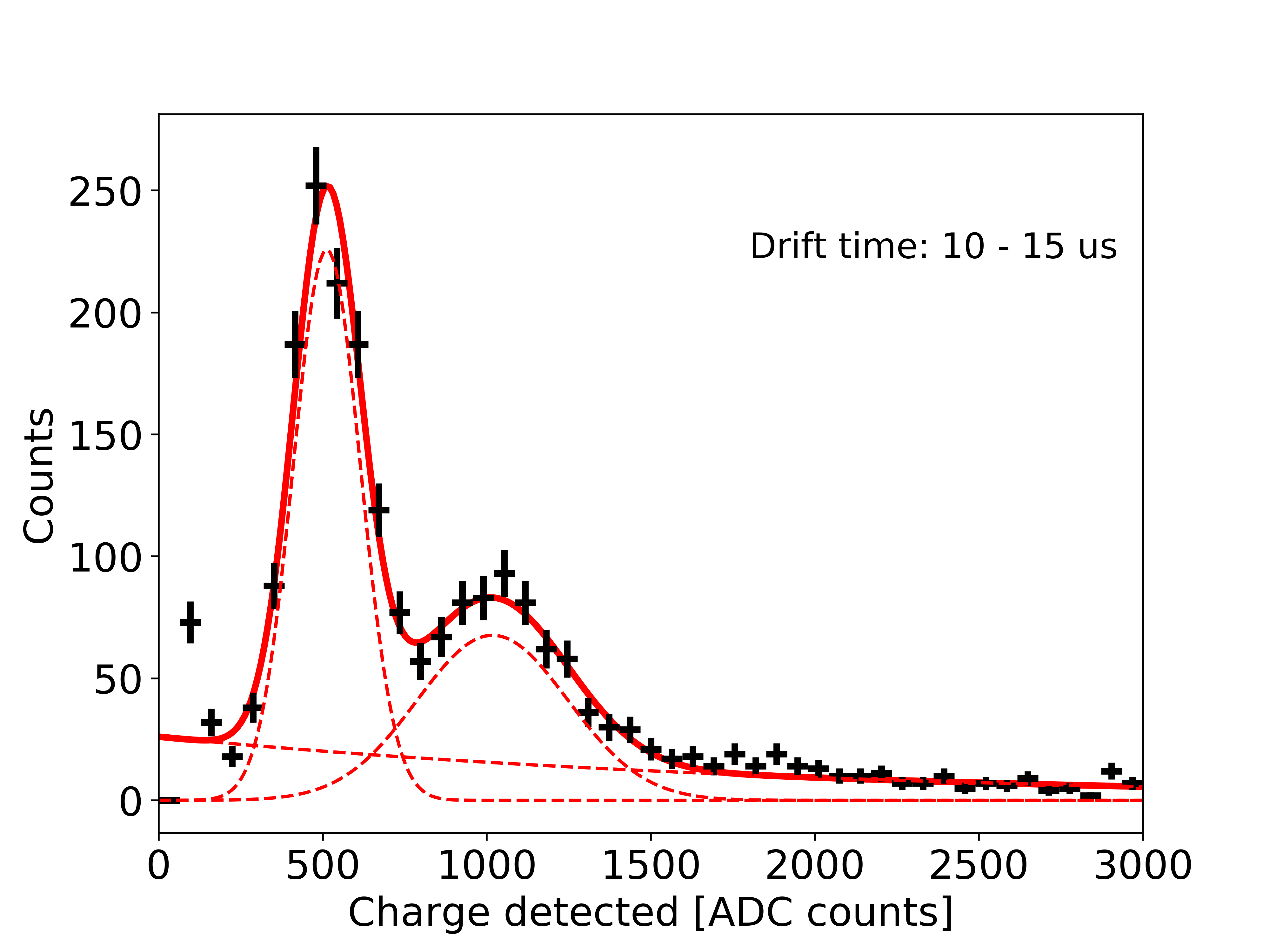}
        \caption{}
        \label{fig:charge_energy_spectrum}
    \end{subfigure}
    \caption{ Measurement of the electron lifetime in the Stanford TPC using \isotope{Xe}{127}. (a) shows the distribution of all events in the charge vs. drift time plane (blue points). The orange and green points show the fitted centroids of the high- and low-energy peaks, respectively. The dashed line shows the best-fit model using Equation~\ref{eq:decay_equation}. (b) shows the distribution for a single slice in drift time. Two peaks corresponding to the 236~keV and 408~keV decay branches are clearly observable.}
\end{figure}

The data are then divided into 5~$\mu$s bins along the drift time axis, as shown in Figure~\ref{fig:charge_vs_drift_time}. An example bin is illustrated by Figure~\ref{fig:charge_energy_spectrum}. The two peaks expected from the 236~keV and 408~keV decay branches are clearly observed. For each drift time bin, the charge energy spectrum is fitted with two Gaussian distributions plus an exponentially-decaying background,
\begin{equation}
    f(x) = a_1\exp\left[-\frac{(x-\mu_1)^2}{2\sigma_1^2}\right] + a_2\exp\left[-\frac{(x-\mu_2)^2}{2\sigma_2^2}\right] + c\exp(-bx),
\label{two_gaussian_function}
\end{equation}
where $a_i$, $\mu_i$ and $\sigma_i$ are the normalization constants, the centroids and the standard deviations, respectively. 

Because the charge sensors are bare strips, the charge induced in the sensors by positive ions must be accounted for. Positive ions, which are produced along with the electrons in an event, induce a surface charge on the strips of opposite sign to the collected charge, leading to an effective reduction in the detected charge relative to the number of electrons collected. On the timescale of a single event, the ions are effectively stationary and their induced charge can be calculated analytically as described in Ref.~\cite{Jewell:2017dzi}. The effective reduction of the detected charge is a function of the distance to the readout strips and the number of strips over which the charge signal is summed (the latter of which affects the surface area over which the ion-induced charge is integrated). This effect is stronger for events closer to the anode, i.e. at shorter drift times, so that its trend is qualitatively opposite to that deriving from the finite electron lifetime. In this analysis, we analytically calculate the strength of the induced charge for a single strip and scale it by the average number of channels above threshold in each event. For the higher-energy (lower-energy) peak, this number is 2.3 (2.0) channels.

\begin{figure}
    \centering
    \includegraphics[width=0.5\textwidth]{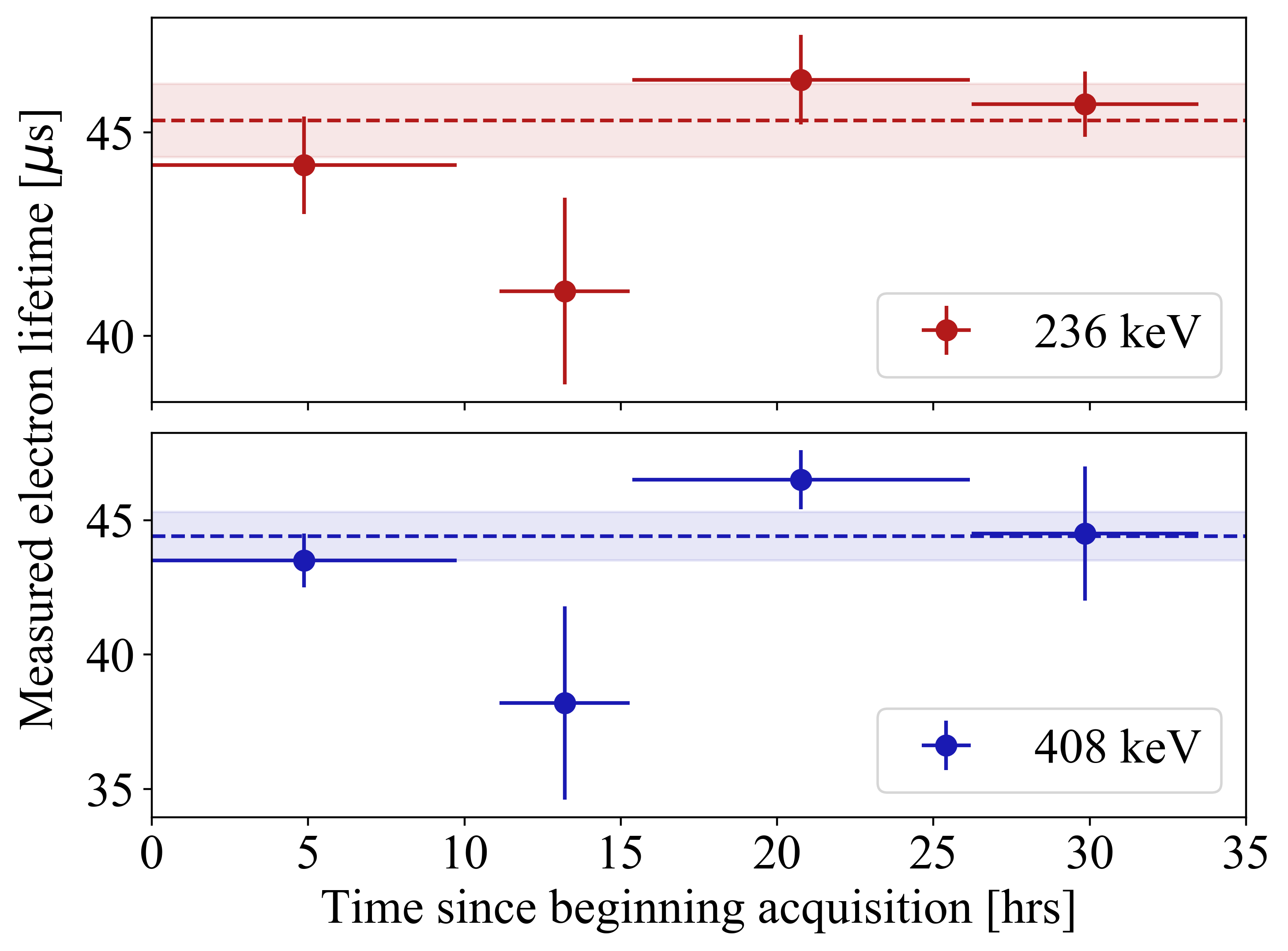}
    \caption{Fitted values of the electron lifetime for both the low-energy (top) and high-energy (bottom) peaks. Data were taken in four separate acquisitions; the $x$ error bars show the duration of the individual acquisition, while the $y$ error bars show the uncertainty in the fitted peak position. The dashed line and error band show the best-fit electron lifetime and uncertainty from the combined dataset. }
    \label{fig:electron_lifetime}
\end{figure}

To extract the electron lifetime, we fit the centroid values for each peak to the function:
\begin{equation}
 Q_e(t) = I(t) \times Q_0\,\,e^{(-t/\tau_e)}
 \label{eq:decay_equation}
\end{equation}
where $I(t)$ is the analytically-calculated positive ion induction signal and $Q_0$ is the initial ionized charge in the event. Both $Q_0$ and $\tau_e$ are left as free parameters. Drift times beyond {50~$\mu$s} are subject to detector threshold effects and are not included in the fit. We fit the data from each of the four acquisitions independently, then, having confirmed their consistency, combine them into a single large dataset which is fitted separately.

The best-fit values of $\tau_e$ are shown in Fig.~\ref{fig:electron_lifetime}, for each dataset individually (data points) and for the full concatenated dataset (dashed line). The concatenated dataset with the best-fit curve from Eq.~\ref{eq:decay_equation} is shown in Figure~\ref{fig:charge_vs_drift_time}. We measure an electron lifetime of approximately \SI{45}{\micro\second}, which is likely limited by outgassing from PVC-insulated wire used in the liquid xenon volume during this particular run. We note that we obtain consistent results for both the 236~keV and 408~keV peaks.

\section{Projections of \isotope{Xe}{127} calibrations in nEXO}
\label{sec:nexo_projections}
To model $^{127}$Xe decay events in nEXO, we use the Geant4-based \texttt{nexo-offline} simulation package, which has been used to estimate nEXO's sensitivity to $0\nu\beta\beta$~\cite{Adhikari_2021}. The software contains a detailed geometry of the nEXO experiment and uses NEST~\cite{NESTZenodo}, tuned to match the measurements in Ref.~\cite{EXO-200:2019bbx}, to model the production of ionization electrons and scintillation photons in the liquid xenon medium. It then models the drift and diffusion of charge through the TPC to the anode plane, properly accounting for the signal development and noise in the charge readout electronics at the level of individual readout channels~\cite{nEXO:2019nye}. In this work we start from the reconstruction algorithm used in Ref.~\cite{Adhikari_2021} and extend it with simplified models of the readout noise to generate the charge and light detection signals.

\subsection{Simulation of $^{127}$Xe calibration datasets}

\label{sec:simulation_of_127xe}
\begin{figure}
    \centering
        \includegraphics[width=0.5\linewidth]{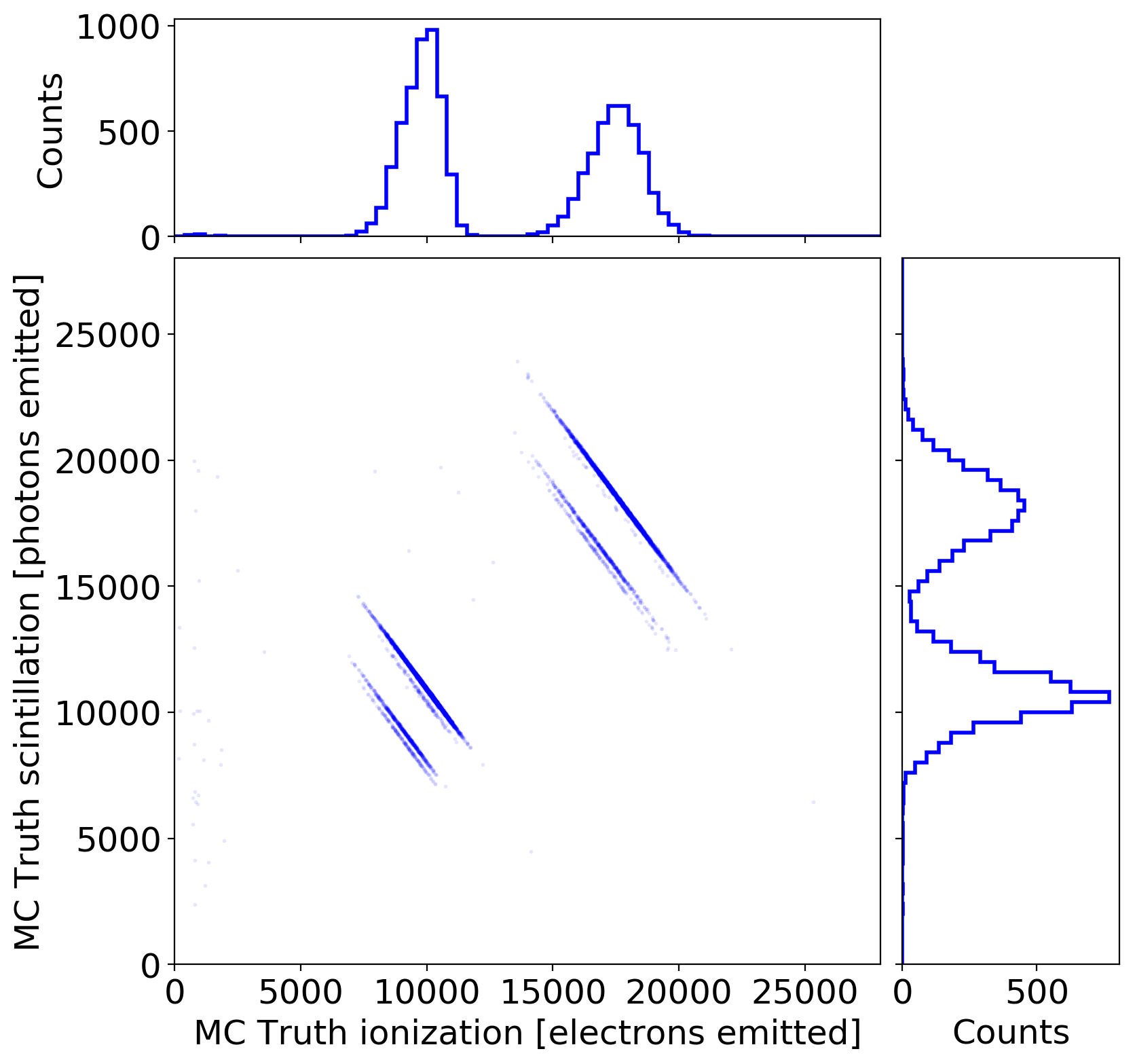}
    \caption{\isotope{Xe}{127} decay events simulated in the active volume of the nEXO TPC. We show the Monte-Carlo-truth distributions of ionization and scintillation produced by the events, as well as their projections onto each observable individually.
    }
    \label{fig:simulated_data_distributions}
\end{figure}

To generate calibration datasets, we simulate $^{127}$Xe decays distributed uniformly throughout the liquid xenon volume in the nEXO TPC. The MC-truth distribution of scintillation light versus charge is shown in Figure~\ref{fig:simulated_data_distributions}, before noise and detection efficiencies are taken into account. As a result of recombination fluctuations, the projections of these distributions into either charge or light appear as two peaks. These two peaks will then be broadened by the detection efficiencies and readout noise of the nEXO detector systems.

In the ionization channel, broadening is introduced both by fluctuations in the charge detection efficiency (due to the finite electron lifetime) and by noise in the readout electronics. The former is included directly in the simulations; the MC-truth $z$ position is used to calculate an expected attenuation probability from Equation~\ref{eq:electron_lifetime}, then the ``detected'' charge is drawn from a binomial distribution to model fluctuations in electron attachment to electronegative impurities in the liquid xenon. The readout noise is modeled for each channel using a normal distribution with a $1\sigma$ width of 600 electrons, which is based on conservative estimates of the noise from simulated waveforms analyzed with a trapezoidal filtering scheme. A similar strategy was used in Ref.~\cite{Adhikari_2021}.

In the scintillation channel, broadening is introduced by fluctuations in the light collection efficiency and the noise due to correlated avalanches in the SiPMs. To model the former, the MC-truth lightmap $\epsilon_{\text{LM}}$ for the nEXO TPC is calculated from a high-statistics light propagation simulation using $5\times 10^8$ point-like photon sources distributed uniformly throughout the TPC. This is performed using Chroma~\cite{chroma}, which uses CUDA-enabled GPUs for fast, high statistics photon transport simulations. The nEXO detector geometry was imported directly into the simulation from CAD software. Details of the optical properties included in the simulation can be found in Ref. \cite{Adhikari_2021}. The resulting position-dependent efficiency values are binned in $r$ and $z$ with a bin size of 0.25~mm by 0.25~mm. A Gaussian blur with a width of 1 bin is then applied to the histogram to produce a smooth and continuous lightmap.

The $xy$-positions of simulated \isotope{Xe}{127} events are determined using an average position of the charge signals for each energy deposition, weighted by the detected charge. The $z$ position is calculated from the drift time and anode position using an electron drift velocity of 0.171~cm/$\mu$s, appropriate for the design field of 400~V/cm~\cite{nEXO_pCDR}. For each event, the number of detected photons is given by
\begin{align}
    n_\mathrm{hit} &= \mathcal{B}\left(S_0,\epsilon_{QE}\cdot\epsilon_{LM}\left(x,y,z\right)\right)\\
    n_\mathrm{av} &= \mathcal{P}(n_\mathrm{hit}\cdot\Lambda)\\
    n_\mathrm{det} &= \frac{n_\mathrm{hit}+n_\mathrm{av}}{\left(1+\Lambda\right)}
\end{align}
where $\mathcal{B}$ and $\mathcal{P}$ represent Binomial and Poisson random variables, respectively, $n_\mathrm{hit}$ is the number of scintillation photons that create a signal in a SiPM, $\epsilon_{LM}(x,y,z)$ is the value of the MC-truth lightmap at the reconstructed event position, $n_\mathrm{av}$ is the number of photons resulting from correlated avalanches, $\Lambda$ is the correlated avalanche fraction, and $n_\mathrm{det}$ is the reconstructed number of detected photons. In this study, we use $\epsilon_{QE}=0.186$ and $\Lambda=0.2$, the current projections for nEXO~\cite{Adhikari_2021}. 
Furthermore, we conservatively assume that the \SI{236}{keV} peak from \isotope{Xe}{127} will fall below the nEXO trigger threshold, and therefore estimate calibration capabilities using only events from the \SI{408}{keV} peak.

\subsection{Electron lifetime calibration}
In a large detector like nEXO, the diffusion of an electron cloud as it drifts across the TPC spreads the charge across more channels, leading to more channels collecting charge for events with longer drift times. This in turn causes more charge to be "lost" below threshold, which results in a systematic bias in the measured charge that mimics a shorter electron lifetime. To avoid this issue, we adopt a "no-threshold" analysis on the charge signals via the following procedure.

First, a temporary 1200~e$^-$ threshold is applied to find channels that collect large amounts of charge. Next, the charge-weighted position of the event is computed using these channels. Finally, the charge is reconstructed by adding the integrated signal on the five nearest channels in each direction (20 in total, counting $\pm \, x$ and $\pm \, y$), regardless of the amount of detected charge on each. As the events tend to be highly localized, this enables us to include the charge collected on channels that do not pass the initial threshold cut, at the cost of introducing additional noise into the reconstructed charge. In addition, the effect of induced charge from the positive ions introduces a bias for events at short drift times similar to that observed in Section~\ref{sec:demonstration}. This effect is exacerbated by the summing of many channels, which increases the induction component to the signal, and leads to a strong rolloff in the reconstructed charge for events closer to the anode. The effect is illustrated by the blue points in Figure~\ref{fig:peak_position_vs_drift}. To correct for this effect, we again use an analytical model of induced charge to calculate the expected bias, making the simplifying assumption that all of the charge is located at the reconstructed charge-weighted position. After applying the correction and performing the "no-threshold" analysis, the biases from diffusion and positive ion induction are removed and the data are well-described by an exponential attenuation, as illustrated by the red points in Figure~\ref{fig:peak_position_vs_drift}. 

\begin{figure}[h]
    \centering
    
    \begin{subfigure}{0.48\linewidth}
        \centering
        \includegraphics[width=0.92\linewidth]{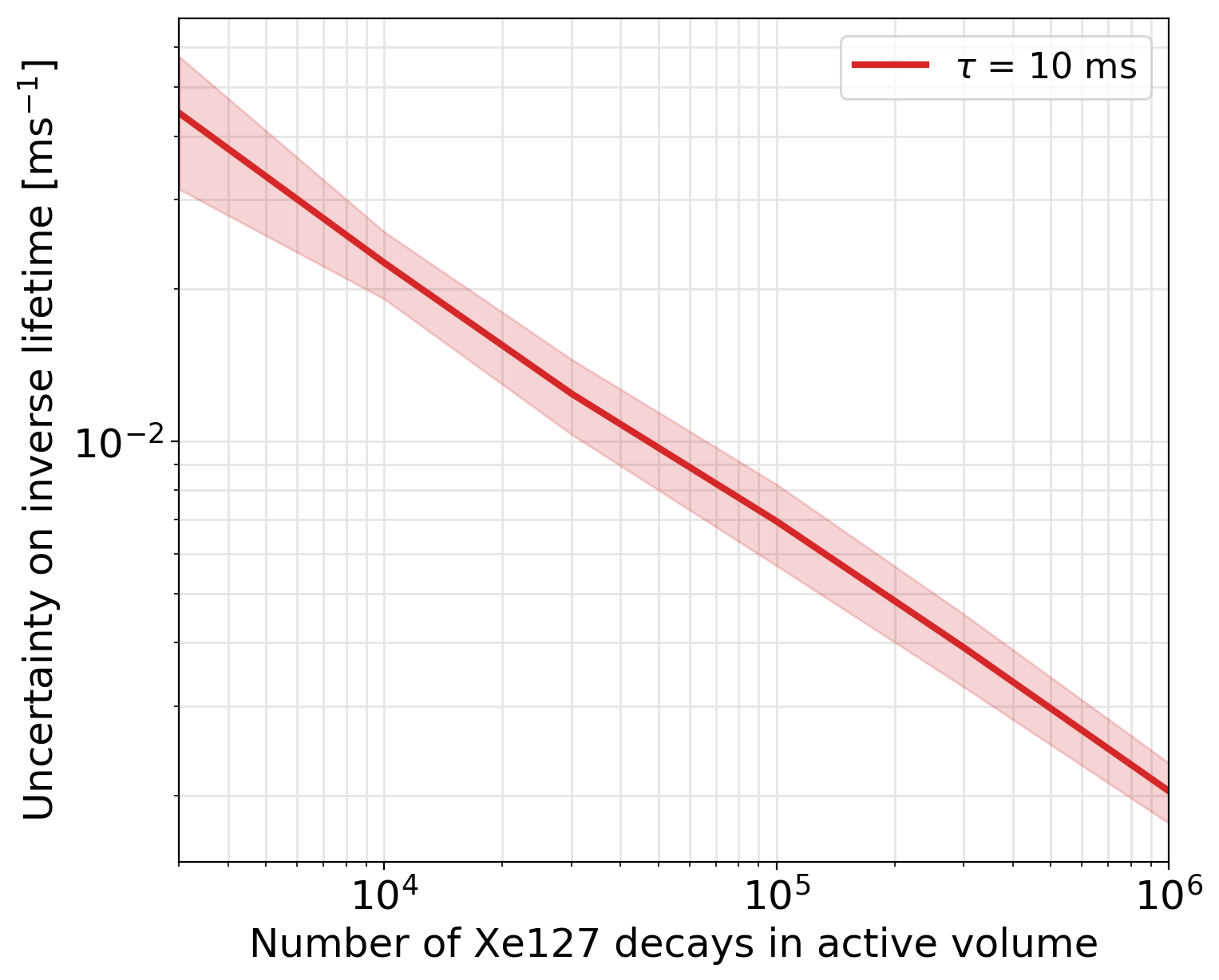}
        \caption{}
        \label{fig:elifetime_uncertainty_vs_num_events}
    \end{subfigure}
    \begin{subfigure}{0.48\linewidth}
        \centering
        \includegraphics[width=0.92\linewidth]{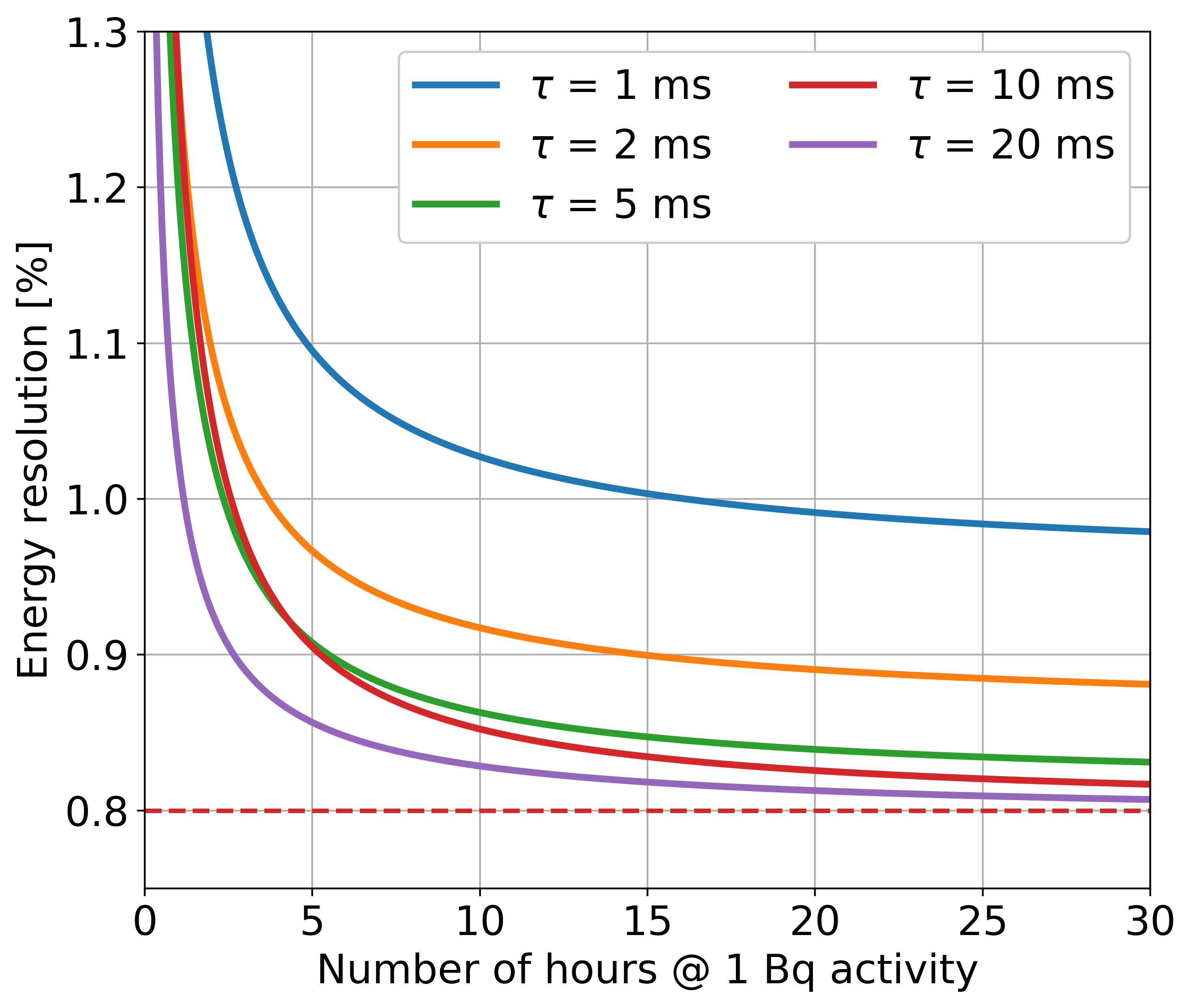}
        \caption{}
        \label{fig:elifetime_resolution}
    \end{subfigure}

    \caption{(a) Uncertainty in the calibration of the inverse electron lifetime $\lambda$ as a function of the number of simulated events in the active volume. The central values and error bands represent the mean and standard deviation of the fitted $\lambda$s over seven simulated datasets. An uncertainty of $10^{-2}$~ms$^{-1}$ corresponds to a relative uncertainty of 10\%. (b) The expected energy resolution of nEXO at $Q_{\beta\beta} = 2.457$~MeV as a function of the integration time, assuming a 1~Bq steady-state activity in the liquid xenon. The expected energy resolution assuming $\tau = 10$~ms and no uncertainty on the electron lifetime is shown by the dashed red line.}
\end{figure}

We evaluate the precision of electron lifetime calibrations as a function of both $\tau_e$ and $N$, where $N$ is the number of events in a given calibration dataset. For each value of $\tau_e$, $10^7$ events are simulated in the active volume of the TPC.
Smaller datasets are created by randomly sampling subsets of these simulations, with replacement. The electron lifetimes considered here range between $\tau_e = $1~--~10~ms, and the sizes of datasets range from $N = 10^3$~--~$10^6$ events. For each ($\tau_e$,$N$) pair, we analyze seven sample datasets; we run the fits and extract $\lambda$ (where $\lambda = 1/\tau_e$) and $\sigma_\lambda$ for each, then take the mean and standard deviation of the seven values to estimate the average performance and the size of expected statistical fluctuations. The result for datasets with $\tau_e = 10$~ms, as a function of $N$, is shown in Figure~\ref{fig:elifetime_uncertainty_vs_num_events}. The average estimated uncertainty, shown by the solid red line, is found to be proportional to $1/\sqrt{N}$, indicating that statistical fluctuations are the dominant source of error in our reconstruction of $\tau_e$. For $\tau_e$ significantly greater than the average drift time (i.e., $\tau_e \geq \sim$1~ms) the absolute uncertainty in $\lambda$ is approximately independent of the true value of $\tau_e$, meaning these conclusions hold for any $\tau_e$ considered here.

\begin{figure}
    \centering
    \begin{subfigure}{0.57\textwidth}
        \includegraphics[width=0.92\linewidth]{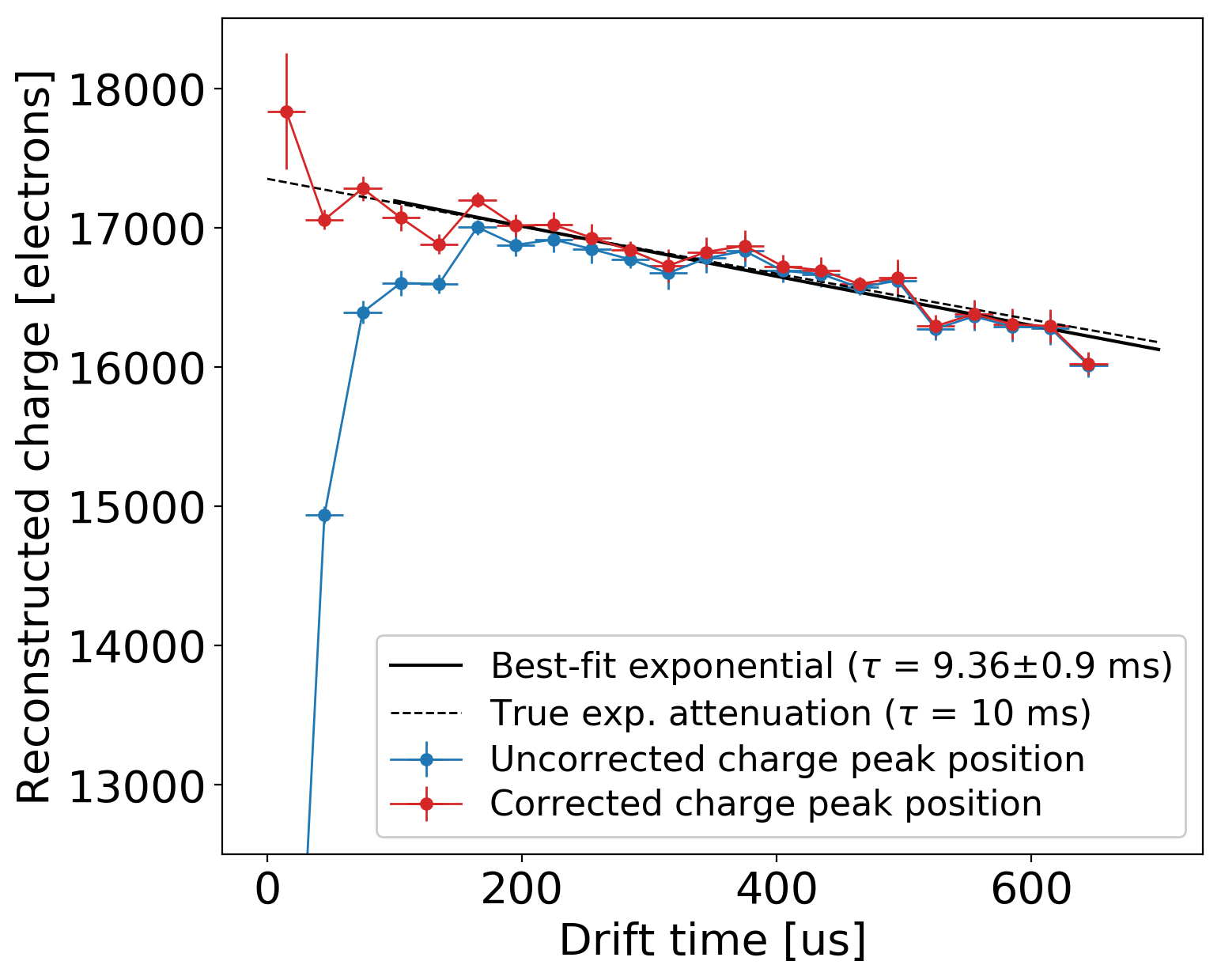}
        \caption{}
        \label{fig:peak_position_vs_drift}
    \end{subfigure}
    
    \caption{(a) Reconstructed charge peak position as a function of the drift time, for a simulated dataset with 30,000 \isotope{Xe}{127} decays in the active volume of the nEXO TPC. The uncorrected data (blue) show the positive ion induction effect on the charge measurement. The data are corrected (red) using the estimated event position as described in the text. The corrected data are fitted to an exponential to determine the electron lifetime.}
    \label{fig:peak_position_vs_drift}
\end{figure}

 Using these results, we then calculate the impact of the electron lifetime calibration uncertainty on the total energy resolution of nEXO. The simulated number of events in the TPC is converted into an integration time, assuming a \isotope{Xe}{127} activity of \SI{1}{Bq} distributed throughout the entire xenon system. The results are shown in Figure~\ref{fig:elifetime_resolution}. When fewer electrons are detected, charge noise makes up a larger fraction of the total charge signal. This results in an asymptotically worse energy resolution at shorter electron lifetimes, despite the independence of the absolute uncertainty in $\lambda$ on the value of $\tau_e$. We find that, for the benchmark case where $\tau_e = 10$~ms, the uncertainty in the electron lifetime calibration after a 24~hr integration period introduces only a 0.03\% absolute broadening of the energy resolution, indicating that a \SI{1}{Bq} source is able to calibrate the electron lifetime on a daily basis with sufficient precision for nEXO.

\subsection{Lightmap calibration}
\begin{figure}
        \centering
    \begin{subfigure}{0.49\linewidth}
        \centering
        \includegraphics[width=\linewidth]{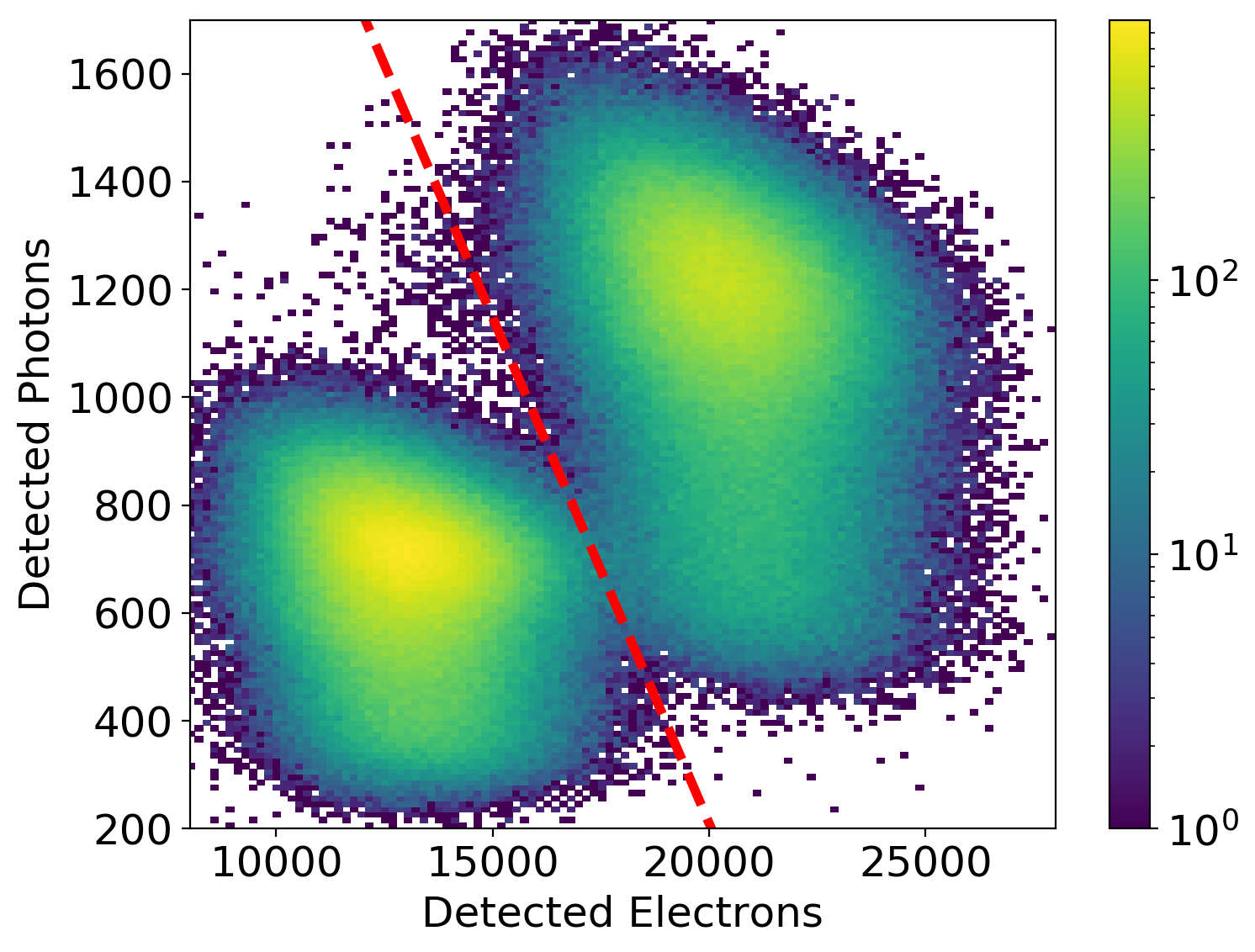}
        \caption{}
        \label{fig:selection_cut}
    \end{subfigure}\hfill
    \begin{subfigure}{0.49\linewidth}
        \centering
        \includegraphics[width=\linewidth]{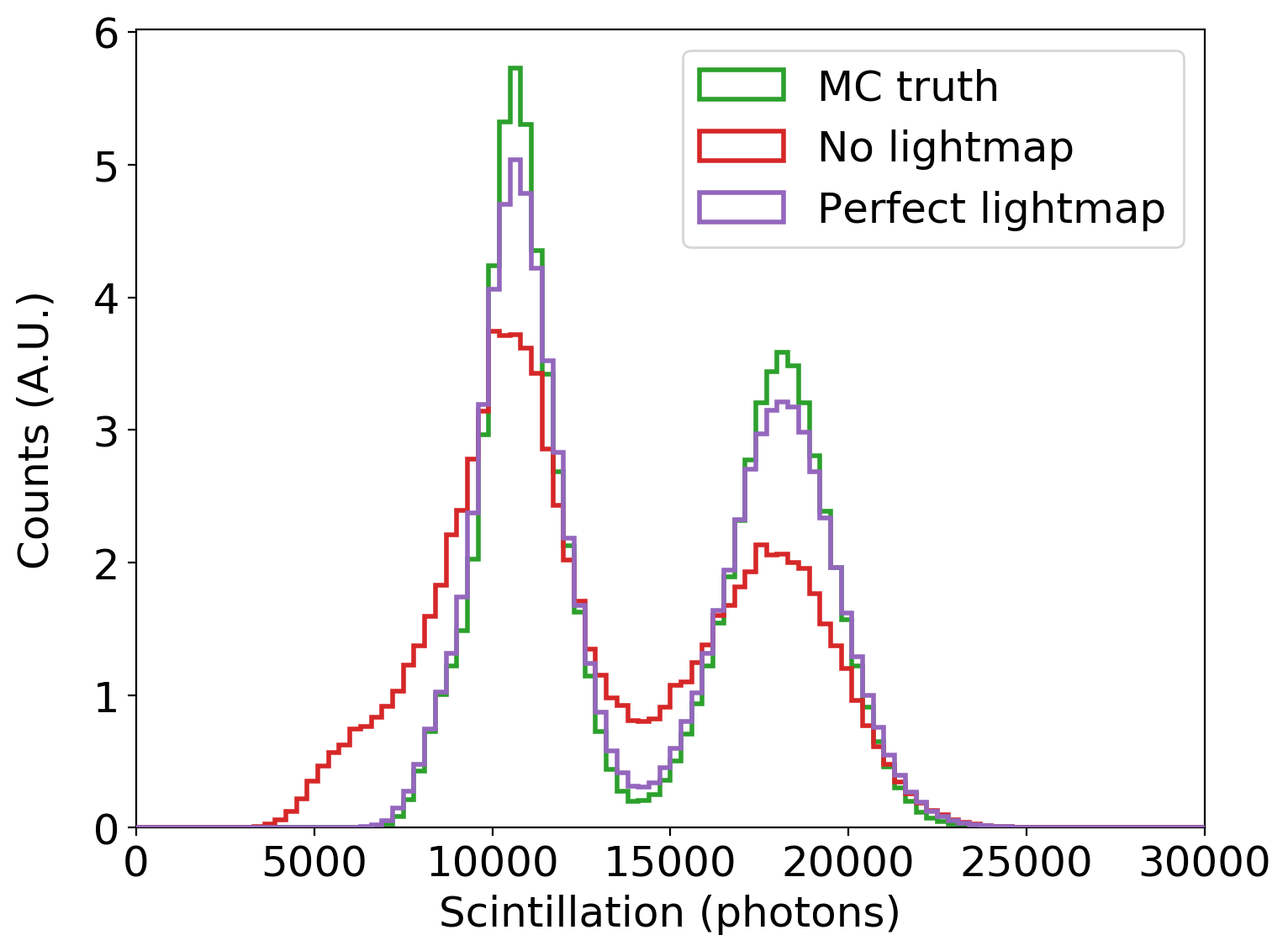}
        \caption{}
        \label{fig:scintillation_photon_hist}
    \end{subfigure}
    \caption{Monte Carlo simulated data from a \isotope{Xe}{127} calibration campaign. 
    In (a), the two peaks are clearly separable with a linear cut in the space of detected electrons and photons. Both peaks appear smeared toward low numbers of scintillation photons as a result of events originating in regions of lower photon transport efficiency. In (b), the effect of a lightmap correction on the scintillation peak width is shown. The green histogram shows the true number of scintillation photons produced, the purple shows how this can be reconstructed assuming the lightmap is calibrated perfectly (i.e. corrected by the same lightmap from which the detected photons were sampled), and the red shows the uncorrected data scaled by a constant.}
    \label{}
\end{figure}

We evaluate the lightmap reconstruction capability by determining the degree to which a lightmap reconstructed from a set of simulated $^{127}$Xe calibration data deviates from the MC-truth lightmap. Figure \ref{fig:selection_cut} shows the uncorrected calibration data, with variations in the photon transport efficiency smearing the light distribution significantly. The photon transport efficiency associated with each calibration event is given by
\begin{equation}
    \epsilon_C=\frac{n_\mathrm{det}}{ \left<S\right> \times \epsilon_{QE}}
\end{equation} where $\left<S\right>=18114$ is the expected number of photons produced for an event corresponding to the 408 keV peak, as determined by NEST. Figure \ref{fig:scintillation_photon_hist} shows the MC-truth scintillation peaks for a sample calibration dataset, along with the scintillation peaks reconstructed both with and without a lightmap correction. From this figure, it should be noted that the width of the uncorrected peaks is dominated by variations in the photon transport efficiency across the detector, while statistical variations in the production and detection of photons contribute less significantly.

There are many techniques that can be used to extract a lightmap from calibration data. The simplest -- choosing an appropriate bin size and binning the data in 3 dimensions -- is not well-suited to nEXO due to the large detector volume. To properly capture the spatial variation in the lightmap, the binning must be sufficiently fine; however this requires a large number of events to avoid empty bins and significant statistical fluctuations between adjacent bins. Techniques that avoid this drawback by interpolating or smoothing over fewer events require the choice of a length parameter based on the data, while the optimal choice of such a parameter often varies by region throughout the TPC. A neural network model avoids these problems. Specifically, it requires no choice of bin size or length scale, it is adaptive to regions in the TPC where the efficiency varies at differing rates, and the computation time scales well with increasing dataset sizes.

\begin{figure}
    \centering
    \includegraphics[width=0.9\linewidth]{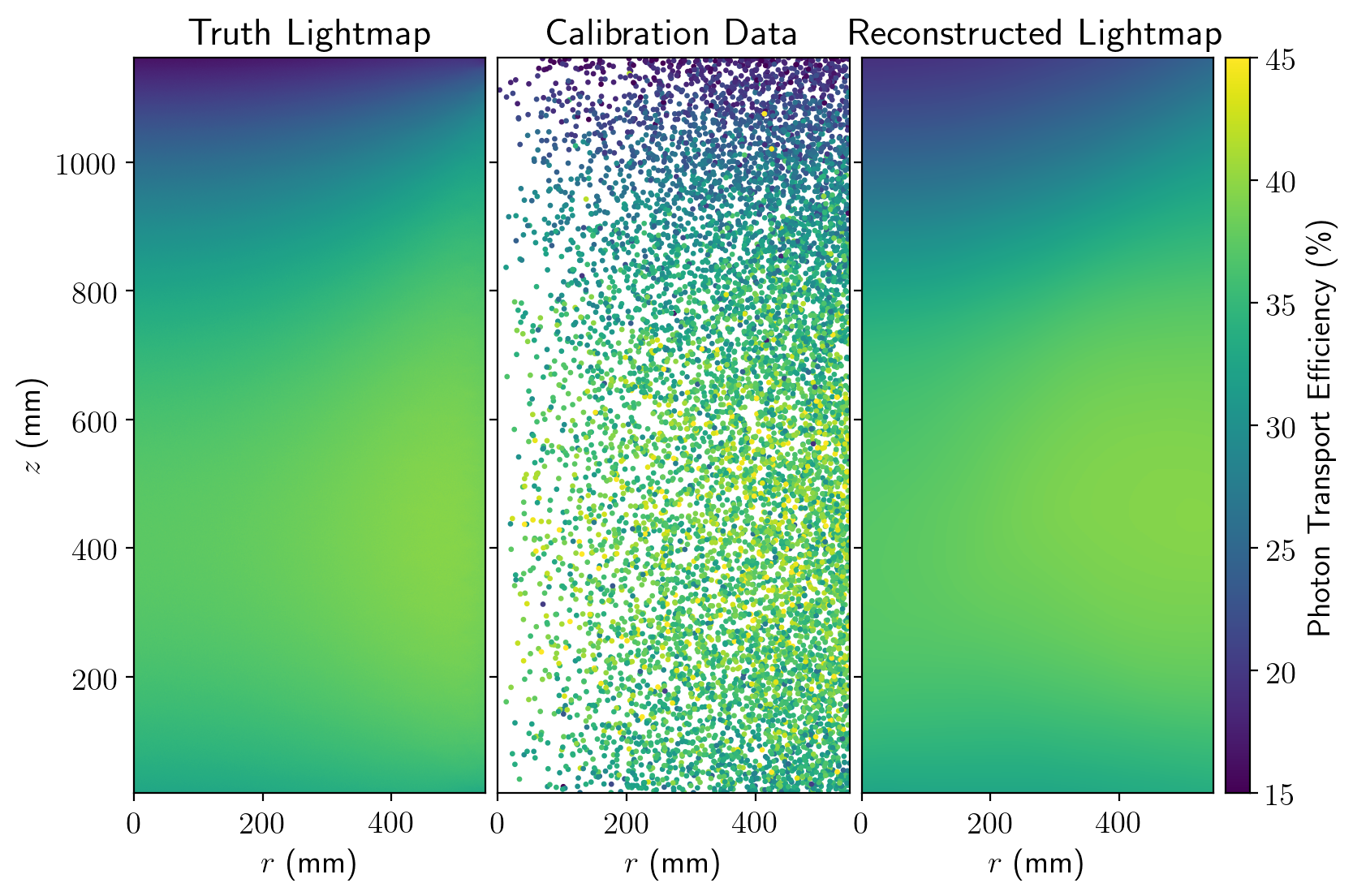}
    \caption{Left: the MC-truth lightmap from which the photon transport efficiency is sampled during the simulation of calibration data. Middle: Points representing the reconstructed positions of $10^4$ calibration events, colored by the photon transport efficiency as calculated from the number of detected photons and the expected number predicted by NEST.  Right: the reconstructed lightmap resulting from training a neural net on the calibration dataset.}    
    \label{fig:sample_calibration}
\end{figure}

The neural net can be trained on the 3-dimensional position coordinates and associated efficiencies of a calibration dataset, giving a continuous efficiency function defined across the full TPC volume. We use a basic feed-forward neural network architecture with five hidden layers. Although the nEXO TPC is designed to be approximately cylindrically symmetric, we reconstruct the lightmap in all three spatial dimensions in order to allow for the possibility of unexpected asymmetries in the real experiment. Neural net hyperparameters were selected to minimize the error in the reconstructed lightmap. Once a satisfactory neural net architecture was chosen, it was trained repeatedly on calibration datasets of varying sizes. The loss function evaluated on both the training and validation datasets was monitored to ensure the neural net was not overfitting to training data. Figure \ref{fig:sample_calibration} shows the full sequence of simulating calibration data from the MC-truth lightmap and then reconstructing the lightmap using the neural net.

Subsets of the simulated events of varying sizes were fed to the neural net to understand the dependence of the reconstructed lightmap accuracy on the number of \isotope{Xe}{127} decays in the active volume. We look specifically at four dataset sizes ranging from $10^3$ to $10^6$ events. For each of these four dataset sizes, 25 calibration datasets were sampled, with replacement, from the full set of $10^7$ events.
\begin{figure}
    \centering
    \includegraphics[width=0.98\linewidth]{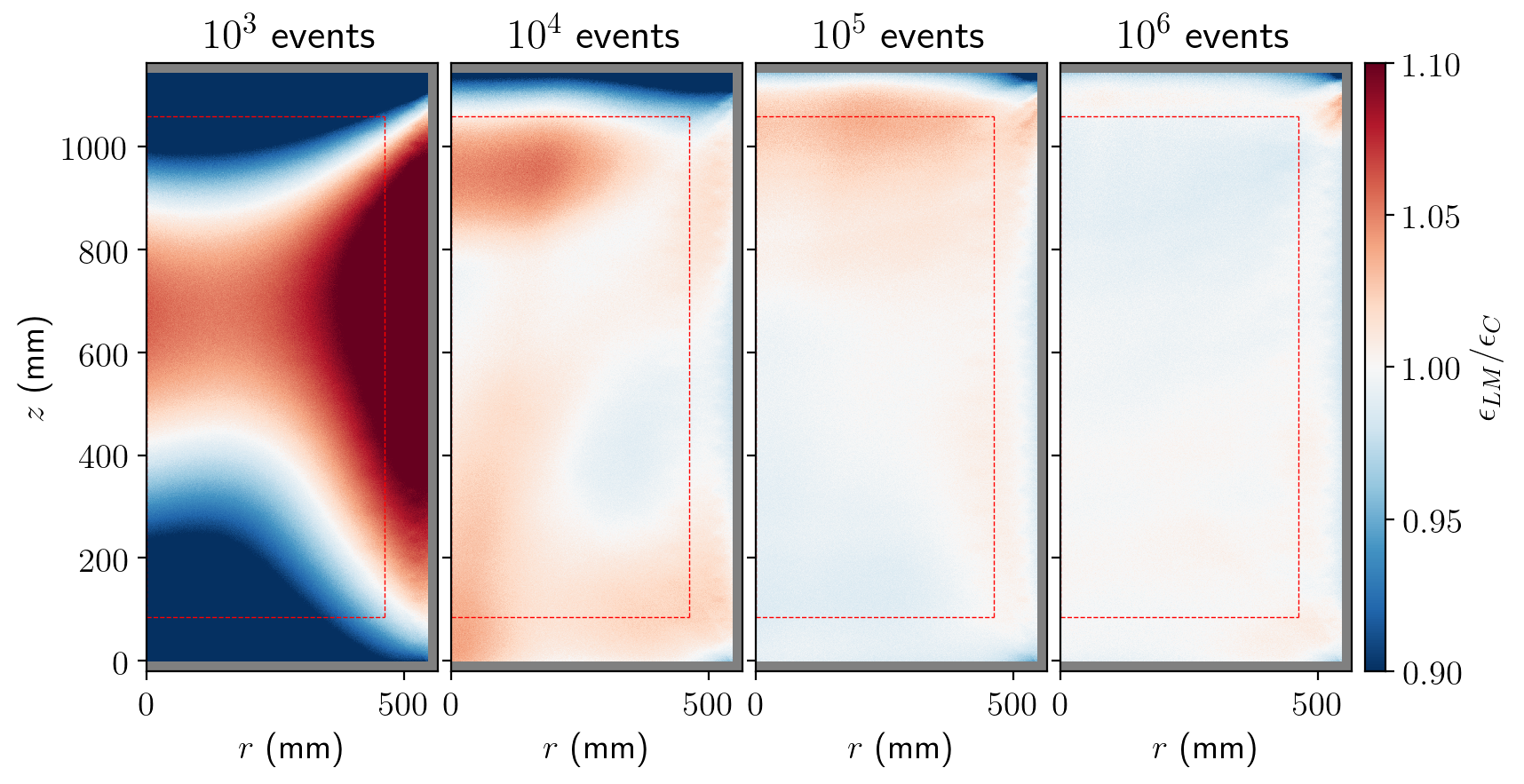}
    \caption{Reconstructed lightmap accuracy at the chosen angular slice of the 3-dimensional lightmap for four sample datasets ranging from $10^3$ to $10^6$ events. The red dotted line shows the boundary that defines the inner 2 tonnes, while the grey band around the detector edges indicates the regions removed  by the 20~mm standoff cut.}    
    \label{fig:lightmap_sample_accuracy}
\end{figure}
To compare the reconstructed lightmap to the MC-truth lightmap, the trained neural net is passed a uniformly-spaced grid of points along an arbitrary angular slice of the TPC. Dividing point-by-point by the corresponding points in the MC-truth lightmap gives the spatially-dependent lightmap reconstruction accuracy in $r$ and $z$ for the chosen slice. This is plotted in Fig. \ref{fig:lightmap_sample_accuracy} for a representative dataset for each of the four dataset sizes. The lightmap error is given by the standard deviation of this accuracy distribution.

\begin{figure}
    \centering
    \includegraphics[width=0.6\linewidth]{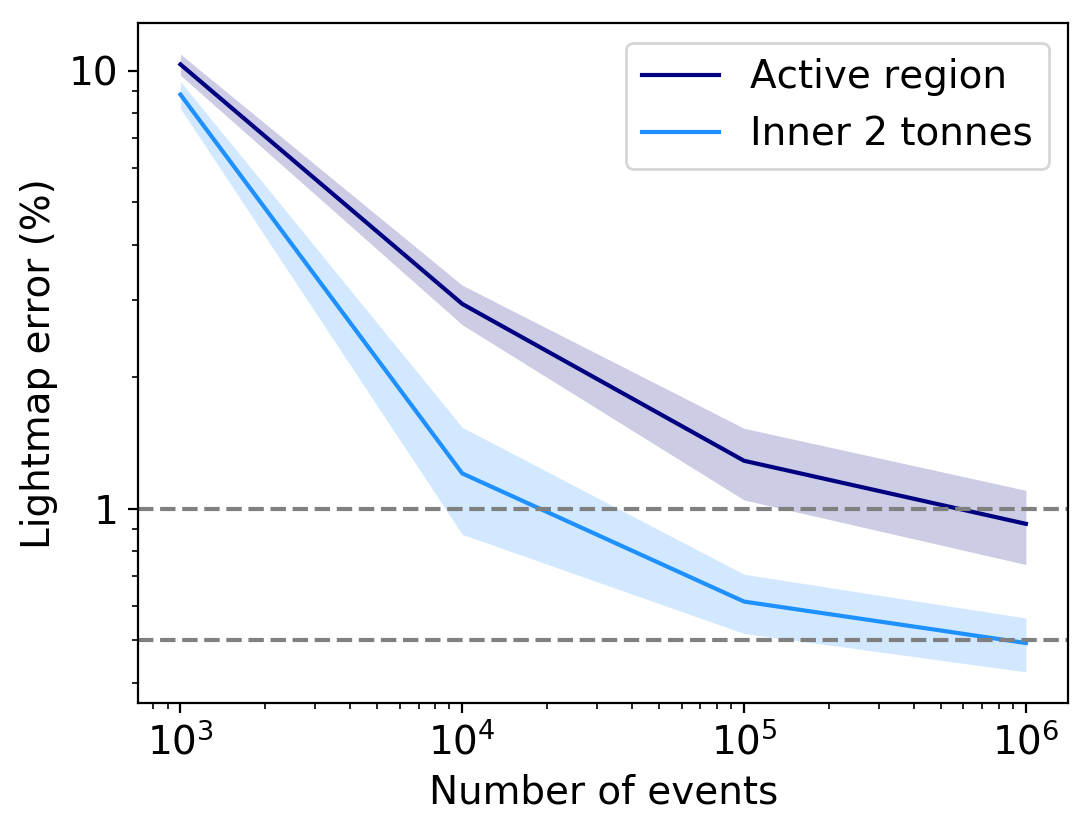}
    \caption{The lightmap error achievable when reconstructing the lightmap from a calibration dataset with a give number of decays in the TPC volume. The solid lines and shaded regions represent the mean and standard deviation of the lightmap errors calculated from all 25 datasets at each point.}
    \label{fig:lightmap_result}
\end{figure}

In the absence of a sufficient density of calibration events to capture the spatial dependence of the true lightmap, the neural net converges to a largely uniform lightmap centered around the average photon transport efficiency. The effect of this overly-uniform lightmap is to introduce a systematic bias in regions with large gradients in the lightmap. With larger calibration datasets, the reconstructed lightmaps converge toward the true lightmap, with the residual errors receding to the edges of the TPC. With increasing statistics, the reduction in residual error continues to the point at which further improvements are limited by both the distance between $\gamma$-ray energy depositions in a single event and the accuracy of the position reconstruction. When the average spacing between calibration events approaches the uncertainty in the event position, further improvements cannot be achieved with larger datasets. Using this lightmap calibration technique, it will be possible to achieve a lightmap error of less than 1\% in the full volume after calibrating for 16 days with a \isotope{Xe}{127} activity maintained at 1 Bq. In the inner two tonnes, this error will fall below 0.5\%. These results are summarized in Figure \ref{fig:lightmap_result}.

\section{Conclusions and prospects for nEXO}
\label{sec:conclusion}
In this work, we have demonstrated the viability of using \isotope{Xe}{127} produced via the neutron activation of \isotope{Xe}{nat} gas as an internal calibration source for liquid xenon TPCs. Such a source was procured, assayed for radioactive contaminants, and used to measure the electron lifetime in a liquid xenon TPC using prototype instrumentation for the nEXO experiment. Simulations show that this source can be used to calibrate both the electron lifetime and the lightmap in nEXO without requiring detector downtime for dedicated calibration campaigns.

\begin{figure}
    \centering
    \includegraphics[width=0.5\textwidth]{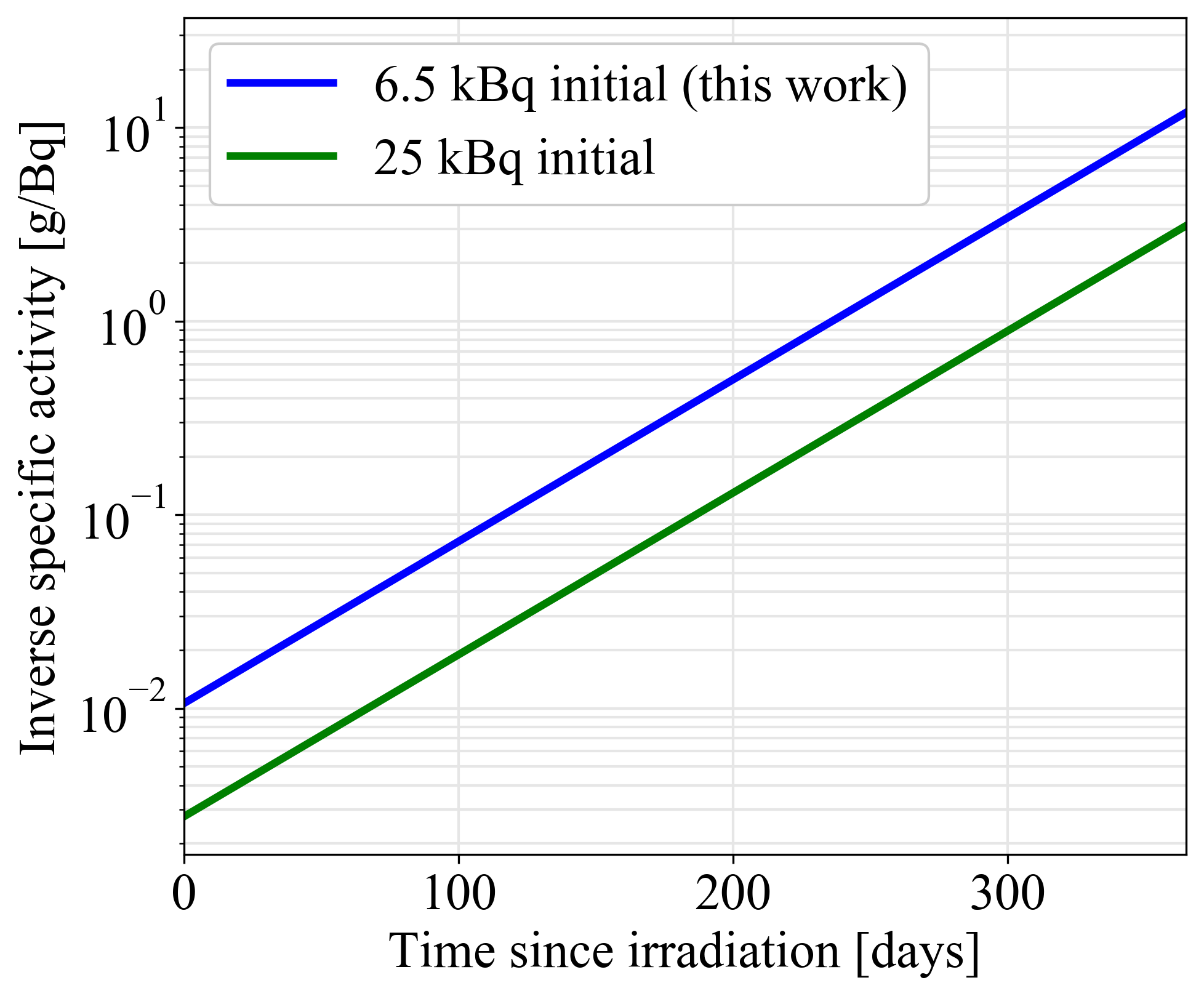}
    \caption{The inverse specific activity for the source used in Section~\ref{sec:demonstration} (blue) and a source with a factor of four higher initial activity (green). In the latter case, a 1~Bq activity corresponds to less than 1~g of gas up to 300~days after activation.}
    \label{fig:inverse_specific_activity}
\end{figure}

In this work we have assumed an activity of 1~Bq is maintained in the liquid xenon continuously. In practice, this could be accomplished by an initial injection of 1~Bq followed every two weeks by injections of $\sim$0.25~Bq to maintain a near-constant activity. Figure~\ref{fig:inverse_specific_activity} shows the amount of xenon gas corresponding to an activity of 1~Bq for two different scenarios: a source with the same activity as the one used in this work, and a source with an activity that is larger by a factor of four. In the latter case, we find that, even up to one year after activation, the source can supply 0.25~Bq injections with less than 1~g of activated xenon gas, resulting in negligible dilution of the enriched xenon in nEXO. Metering the injected gas could be accomplished by a simple volume-sharing scheme such as that used in Section~\ref{sec:demonstration}, but with a larger ratio of initial volume to expansion volume to permit higher precision.

\acknowledgments

The authors thank Harrison Redman of the Health Physics Group at Stanford for assistance with HPGe radioassay measurements, and Wesley Frey at MNRC for help with source development. The authors gratefully acknowledge support for nEXO from the Office of Nuclear Physics within DOE's Office of Science, and NSF in the United States; from the National Research Foundataion (NRF) of South Africa; from NSERC, CFI, FRQNT, NRC, and the McDonald Institute (CFREF) in Canada; from IBS in Korea; from RFBR in Russia; and from CAS and NSFC in China.

\providecommand{\href}[2]{#2}\begingroup\raggedright\endgroup




\end{document}